\documentclass{nature3}
\usepackage{graphicx}
\usepackage{float}
\usepackage{verbatim}
\usepackage{hyperref}
\usepackage{amsmath}
\usepackage{amssymb}
\usepackage{aas_macros_nature}
\usepackage{xfrac}

\usepackage{lineno}

\usepackage{xcolor}
\usepackage[flushleft]{threeparttable}

\definecolor{my_color}{HTML}{3a18b1}

\definecolor{new_color}{HTML}{CF0000}% this is a maroon
\definecolor{new_black}{HTML}{000000}% this is a maroon

\newcommand*{\bedit}{\textcolor{new_black}}
\newcommand*{\bedittwo}{\textcolor{new_black}}

\newcommand{\rsun}{\ensuremath{R_\odot}}
\newcommand{\msun}{\ensuremath{M_\odot}}

\newcommand{\arcsecond}{\ensuremath{^{\prime\prime}}}

\newcommand{\farcm}{\mbox{\ensuremath{.\mkern-4mu^\prime}}}%    % fractional arcminute symbol: 0.'0
\newcommand{\farcs}{\mbox{\ensuremath{.\!\!^{\prime\prime}}}}%  % fractional arcsecond symbol: 0.''0

\newcommand{\halpha}{H$\alpha$}

\newcommand{\rearth}{\ensuremath{R_\oplus}}

\newcommand{\mjup}{\ensuremath{M_{\rm J}}}

\newcommand{\microjanskys}{\ensuremath{\mathbf{{\rm \mu Jy}}}}

\newcommand{\project}[1]{\textsl{#1}}                               
\newcommand{\JWST}{\project{JWST}}      
\newcommand{\jameswebb}{\project{James Webb}}                               

\newcommand{\HST}{\project{HST}} 
\newcommand{\PLATO}{\project{PLATO}} 

\newcommand{\Spitzer}{\project{Spitzer}}

\newcommand{\Gaia}{\project{Gaia}}

\newcommand{\TESS}{\project{TESS}}

\newcommand{\kms}{\ensuremath{\rm km\,s^{-1}}}

\newcommand{\gcc}{\ensuremath{\rm g\,cm^{-3}}}

\newcommand{\rprstb}{7.28}
\newcommand{\urprstb}{\ensuremath{\pm}0.65}
\newcommand{\tdurb}{7.998}
\newcommand{\utdurb}{\ensuremath{\pm}0.023}
\newcommand{\arstb}{336}
\newcommand{\uarstb}{\ensuremath{\pm}14}
\newcommand{\inclb}{88.778}
\newcommand{\uinclb}{\ensuremath{\pm}0.059}
\newcommand{\impb}{7.16}
\newcommand{\uimpb}{\ensuremath{\pm}0.65}
\newcommand{\rplb}{10.4}
\newcommand{\urplb}{\ensuremath{\pm}1.0}
\newcommand{\perplb}{1.4079405}
\newcommand{\uperplb}{\ensuremath{\pm}0.0000011}
\newcommand{\ttransitb}{2458779.3750828}
\newcommand{\uttransitb}{\ensuremath{\pm}0.0000034}
\newcommand{\ab}{0.0204}
\newcommand{\uab}{\ensuremath{\pm}0.0012}
\newcommand{\teqb}{163}
\newcommand{\uteqb}{\ensuremath{^{+14}_{-18}}}
\newcommand{\fluxb}{0.181}
\newcommand{\ufluxb}{\ensuremath{\pm}0.018}
\newcommand{\dspit}{0.004}
\newcommand{\udspit}{\ensuremath{\pm}0.029}

\newcommand{\rprstbecc}{10.8}
\newcommand{\urprstbecc}{\ensuremath{^{+3.9}_{-2.6}}}
\newcommand{\tdurbecc}{7.945}
\newcommand{\utdurbecc}{\ensuremath{\pm}0.037}
\newcommand{\arstbecc}{325}
\newcommand{\uarstbecc}{\ensuremath{\pm}18}
\newcommand{\inclbecc}{87.4}
\newcommand{\uinclbecc}{\ensuremath{^{+1.0}_{-1.7}}}
\newcommand{\impbecc}{10.7}
\newcommand{\uimpbecc}{\ensuremath{^{+3.9}_{-2.6}}}
\newcommand{\rplbecc}{15.4}
\newcommand{\urplbecc}{\ensuremath{^{+5.5}_{-3.7}}}
\newcommand{\perplbecc}{1.4079405}
\newcommand{\uperplbecc}{\ensuremath{\pm}0.0000011}
\newcommand{\ttransitbecc}{2458779.37508}
\newcommand{\uttransitbecc}{\ensuremath{\pm}0.00012}
\newcommand{\abecc}{0.0198}
\newcommand{\uabecc}{\ensuremath{\pm}0.0014}
\newcommand{\teqbecc}{164}
\newcommand{\uteqbecc}{\ensuremath{^{+14}_{-18}}}
\newcommand{\fluxbecc}{0.212}
\newcommand{\ufluxbecc}{\ensuremath{^{+0.041}_{-0.029}}}
\newcommand{\eccb}{ \ensuremath{<} 0.68}
\newcommand{\dspitecc}{0.004}
\newcommand{\udspitecc}{\ensuremath{\pm}0.028}

\newcommand{\appropto}{\mathrel{\vcenter{
  \offinterlineskip\halign{\hfil$##$\cr
    \propto\cr\noalign{\kern2pt}\sim\cr\noalign{\kern-2pt}}}}}

\title{A Giant Planet Candidate Transiting a White Dwarf}

\author{Andrew Vanderburg$^{1,2,3}$,
Saul A. Rappaport$^{4}$,
Siyi Xu$^{5}$,
Ian Crossfield$^{6}$,
Juliette C. Becker$^{7,8}$,
Bruce Gary$^{9}$,
Felipe Murgas$^{10,11}$,
Simon Blouin$^{12}$,
Thomas G. Kaye$^{13,14}$,
Enric Palle$^{10,11}$,
Carl Melis$^{15}$,
Brett M. Morris$^{16}$,
Laura Kreidberg$^{17,18}$,
Varoujan Gorjian$^{19}$,
Caroline V. Morley$^{2}$,
Andrew W. Mann$^{20}$,
Hannu Parviainen$^{10,11}$,
Logan A. Pearce$^{21,22}$,
Elisabeth R. Newton$^{23}$,
Andreia Carrillo$^{2,24}$,
Ben Zuckerman$^{25}$,
Lorne Nelson$^{26}$,
Greg Zeimann$^{27}$,
Warren R. Brown$^{18}$,
Ren\'e Tronsgaard$^{28}$,
Beth Klein$^{25}$,
George R. Ricker$^{4}$,
Roland K. Vanderspek$^{4}$,
David W. Latham$^{18}$,
Sara Seager$^{4,29,30}$,
Joshua N. Winn$^{31}$,
Jon M. Jenkins$^{32}$,
Fred C. Adams$^{33,34}$,
Bj\"orn Benneke$^{35,36}$,
David Berardo$^{4}$,
Lars A. Buchhave$^{28}$,
Douglas A. Caldwell$^{37,32}$,
Jessie L. Christiansen$^{38}$,
Karen A. Collins$^{18}$,
Knicole D. Col\'{o}n$^{39}$,
Tansu Daylan$^{4,40}$,
John Doty$^{41}$,
Alexandra E. Doyle$^{42}$,
Diana Dragomir$^{43}$,
Courtney Dressing$^{44}$,
Patrick Dufour$^{35,36}$,
Akihiko Fukui$^{45,10}$,
Ana Glidden$^{4,29}$,
Natalia M. Guerrero$^{4}$,
Xueying Guo$^{4}$,
Kevin Heng$^{16}$,
Andreea I. Henriksen$^{28}$,
Chelsea X. Huang$^{4,46}$,
Lisa Kaltenegger$^{47,48}$,
Stephen R. Kane$^{49}$,
John A. Lewis$^{18}$,
Jack J. Lissauer$^{32}$,
Farisa Morales$^{19,50}$,
Norio Narita$^{51,52,53,10,54}$,
Joshua Pepper$^{55}$,
Mark E. Rose$^{32}$,
Jeffrey C. Smith$^{37,32}$,
Keivan G. Stassun$^{56,57}$,
Liang Yu$^{58,4}$
}

\begin{document}

\maketitle
\scriptsize
\begin{affiliations}
\item Department of Astronomy, University of Wisconsin-Madison, Madison, WI 53706, USA
\item Department of Astronomy, The University of Texas at Austin, Austin, TX 78712, USA
\item NASA Sagan Fellow
\item Department of Physics and Kavli Institute for Astrophysics and Space Research, Massachusetts Institute of Technology, Cambridge, MA 02139, USA
\item NSF’s NOIRLab/Gemini Observatory, 670 N. A’ohoku Place, Hilo, Hawaii, 96720, USA
\item Dept. of Physics and Astronomy, University of Kansas, 1251 Wescoe Hall Dr., Lawrence, KS 66045, USA
\item Division of Geological and Planetary Sciences, California Institute of Technology, Pasadena, CA 91125, USA
\item 51 Pegasi b Fellow
\item Hereford Arizona Observatory, Hereford, AZ 85615, USA
\item Instituto de Astrof\'isica de Canarias (IAC), E-38200 La Laguna, Tenerife, Spain
\item Dept. Astrof\'isica, Universidad de La Laguna (ULL), E-38206 La Laguna, Tenerife, Spain
\item Los Alamos National Laboratory, P.O. Box 1663, Mail Stop P365, Los Alamos, NM 87545, USA
\item Raemor Vista Observatory, 7023 E. Alhambra Dr., Sierra Vista, AZ 85650, USA
\item Laboratory for Space Research, The University of Hong Kong, Pokfulam, Hong Kong, China
\item Center for Astrophysics and Space Sciences, UCSD, CA 92093-0424, USA
\item Center for Space and Habitability, University of Bern, Gesellschaftsstrasse 6, CH-3012, Bern, Switzerland
\item Max Planck Institute for Astronomy, K{\"o}nigstuhl 17, 69117 Heidelberg, Germany
\item Center for Astrophysics $|$  Harvard \& Smithsonian, Cambridge, MA 02138, USA
\item NASA Jet Propulsion Laboratory, 4800 Oak Grove Dr, Pasadena, CA 91109, USA
\item Department of Physics and Astronomy, University of North Carolina at Chapel Hill, Chapel Hill, NC 27599, USA
\item Steward Observatory, University of Arizona, Tucson, AZ 85721, USA
\item NSF Graduate Research Fellow
\item Department of Physics and Astronomy, Dartmouth College, Hanover, NH 03755, USA
\item Large Synoptic Survey Telescope Corporation Data Science Fellow
\item Department of Physics and Astronomy, University of California, Los Angeles, CA 90095-1562, USA
\item Department of Physics and Astronomy, Bishop’s University, Sherbrooke, QC J1M 1Z7, Canada
\item Hobby Eberly Telescope, University of Texas, Austin, Austin, TX, 78712, USA
\item DTU Space, National Space Institute, Technical University of Denmark, Elektrovej 328, DK-2800 Kgs. Lyngby, Denmark
\item Department of Earth and Planetary Sciences, MIT, 77 Massachusetts Avenue, Cambridge, MA 02139, USA
\item Department of Aeronautics and Astronautics, MIT, 77 Massachusetts Avenue, Cambridge, MA 02139, USA
\item Department of Astrophysical Sciences, Princeton University, 4 Ivy Lane, Princeton, NJ 08544, USA
\item NASA Ames Research Center, Moffett Field, CA, 94035, USA
\item Physics Department, University of Michigan, Ann Arbor, MI 48109, USA
\item Astronomy Department, University of Michigan, Ann Arbor, MI 48109, USA
\item D\'epartment de Physique, Universit\'e de Montr\'eal, Montr\'eal, QC H3C 3J7, Canada
\item Institut de Recherche sur les Exoplan\`etes (iREx), Universit\'e de Montr\'eal, Montr\'eal, QC H3C 3J7, Canada
\item SETI Institute, Mountain View, CA 94043, USA
\item Caltech/IPAC-NASA Exoplanet Science Institute, Pasadena, CA 91106, USA
\item NASA Goddard Space Flight Center, Exoplanets and Stellar Astrophysics Laboratory (Code 667), Greenbelt, MD 20771, USA
\item Kavli Fellow
\item Noqsi Aerospace, Ltd., 15 Blanchard Avenue, Billerica, MA 01821, USA
\item Department of Earth, Planetary, and Space Sciences, University of California, Los Angeles, Los Angeles, CA, USA
\item Department of Physics and Astronomy, University of New Mexico, Albuquerque, NM, USA
\item Department of Astronomy, University of California - Berkeley, Berkeley, CA, 94720, USA
\item Department of Earth and Planetary Science, Graduate School of Science, The University of Tokyo, 7-3-1 Hongo, Bunkyo-ku, Tokyo 113-0033, Japan
\item Juan Carlos Torres Fellow
\item Cornell University, Astronomy and Space Sciences Building, Ithaca, NY 14850, USA
\item Carl Sagan Institute, Space Science Building 311, Ithaca, NY 14850, USA
\item Department of Earth and Planetary Sciences, University of California, Riverside, CA 92521, USA
\item Department of Physics and Astronomy, Moorpark College, 7075 Campus Road, Moorpark, CA 93021, USA
\item Astrobiology Center, 2-21-1 Osawa, Mitaka, Tokyo 181-8588, Japan
\item JST, PRESTO, 2-21-1 Osawa, Mitaka, Tokyo 181-8588, Japan
\item National Astronomical Observatory of Japan, 2-21-1 Osawa, Mitaka, Tokyo 181-8588, Japan
\item Komaba Institute for Science, The University of Tokyo, 3-8-1 Komaba, Meguro, Tokyo 153-8902, Japan
\item Department of Physics, Lehigh University, 16 Memorial Drive East, Bethlehem, PA 18015, USA
\item Department of Physics and Astronomy, Vanderbilt University, Nashville, TN 37235 USA
\item Department of Physics, Fisk University, Nashville, TN 37208, USA
\item ExxonMobil Upstream Integrated Solutions, Spring, TX 77389, USA
\end{affiliations}
\normalsize

\begin{abstract}
Astronomers have discovered thousands of planets outside the solar system\cite{akeson2013}, most of which orbit stars that will eventually evolve into red giants and then into white dwarfs. During the red giant phase, any close-orbiting planets will be engulfed by the star\cite{Villaver2009}, but more distant planets can survive this phase and remain in orbit around the white dwarf\cite{luhman2011,marsh2014}. Some white dwarfs show evidence for rocky material floating in their atmospheres\cite{jura2003}, in warm debris disks\cite{kilic2005, becklin2005, gansicke2006, wilson2019}, or orbiting very closely\cite{wd1145, manser2019, vanderbosch2019}, which has been interpreted as the debris of rocky planets that were scattered inward and tidally disrupted\cite{debes2002}. Recently, the discovery of a gaseous debris disk with a composition similar to ice giant planets\cite{gansicke2019} demonstrated that massive planets might also find their way into tight orbits around white dwarfs, but it is unclear whether the planets can survive the journey. So far, the detection of intact planets in close orbits around white dwarfs has remained elusive. Here, we report the discovery of a giant planet candidate transiting the white dwarf WD\,1856+534 (TIC 267574918) every 1.4 days. The planet candidate is roughly the same size as Jupiter and is no more than 14 times as massive (with 95\% confidence). Other cases of white dwarfs with close brown dwarf or stellar companions are explained as the consequence of common-envelope evolution, wherein the original orbit is enveloped during the red-giant phase and shrinks due to friction. In this case, though, the low mass and relatively long orbital period of the planet candidate make \bedit{common-envelope evolution less likely}. Instead, the WD\,1856+534 system seems to demonstrate that giant planets can be scattered into tight orbits without being tidally disrupted, and motivates searches for smaller transiting planets around white dwarfs.

\end{abstract}

WD\,1856+534 (hereafter, WD\,1856 for brevity) is located 25 parsecs away in a visual triple star system. It has an effective temperature of $4710\pm 60$~Kelvin and became a white dwarf 5.9 $\pm$ 0.5  billion years ago, based on theoretical models for how white dwarfs cool over time. The total system age, including the star's main sequence lifetime, must be older. Table \ref{parameters} gives the other key parameters of the star. WD 1856 is one of thousands of white dwarfs that was targeted for observations with NASA's Transiting Exoplanet Survey Satellite (\TESS), in order to search for any periodic dimming events caused by planetary transits. A statistically significant transit-like event was detected by the \TESS\ Science Processing Operations Center (SPOC) pipeline based on 28 days of data acquired between 18 July and 14 August, 2019. The signal was rejected by an automated classification system designed to identify planets around main-sequence stars. We noticed the signal in a visual inspection of all possible transit-like events detected around white dwarfs.
As usual, caution is required when interpreting \TESS\ data because of the relatively coarse angular resolution; in this case, the white dwarf was blended together with several much brighter stars in the \TESS\ images.
However, the signal's duration of $\approx$8 minutes is much shorter than the usual duration of $\gtrsim$30 minutes for the transit of a main-sequence star, strongly suggesting that the transit signal originates from the white dwarf and not the other stars. 

\begin{figure}[t!]
\centering
\includegraphics[width=0.99\textwidth]{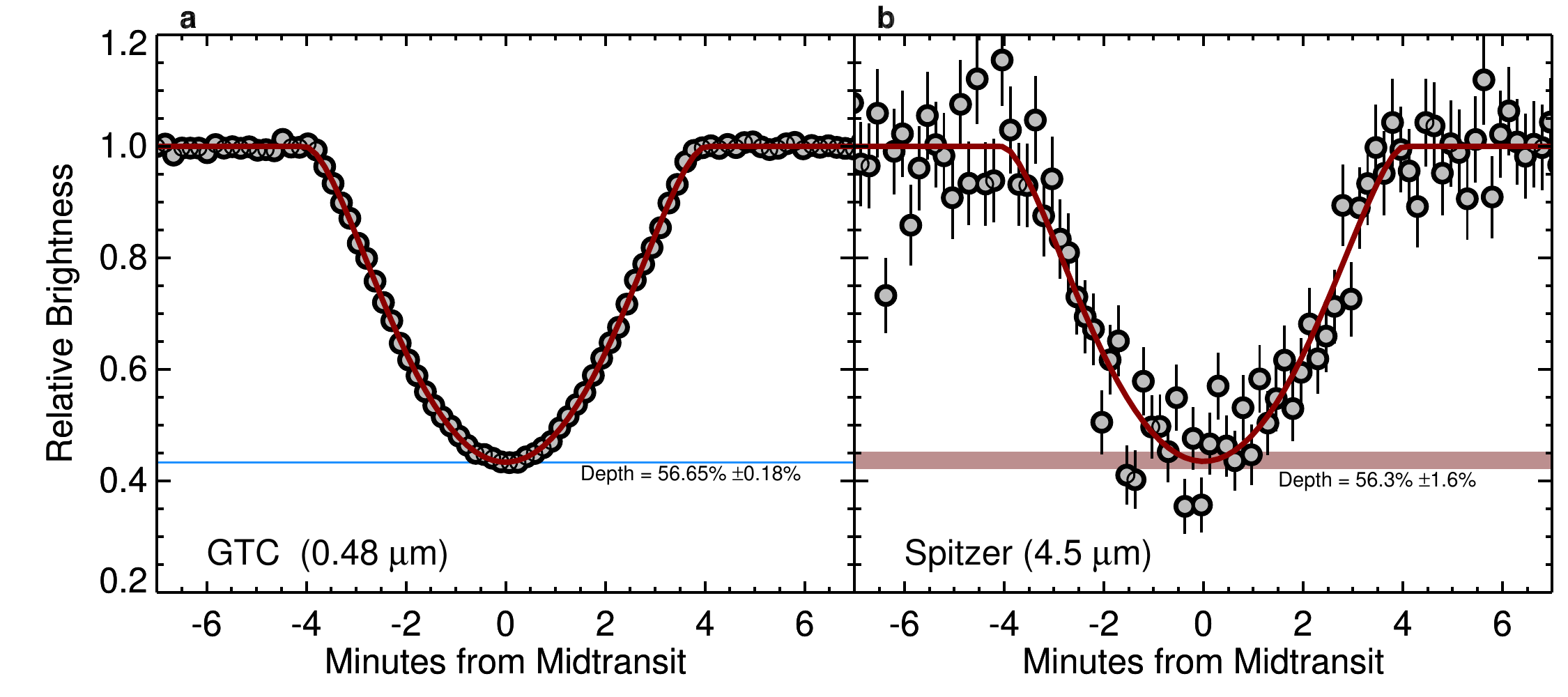}
\caption{\bedittwo{Transit observations of WD 1856.} \textbf{a}, Optical transit observations with the Gran Telescopio Canarias and \textbf{b},
infrared transit observations with the \Spitzer\ Space Telescope.
The red curves are the best-fitting models. The horizontal colored shaded regions (light blue for GTC, light red for \Spitzer) show the 68\% confidence interval for the maximum loss of light. Any thermal emission from the transiting body would have
led to a smaller loss of light at infrared wavelengths. The lack of any
observed difference implies that the transiting body has a mass smaller than
13.8 Jupiter masses (with 95\% confidence). \bedittwo{Each \Spitzer\ point is an average of five exposures (each with a two second exposure time), and the error bars show the 1$\sigma$ error on the mean.} The uncertainties on the GTC points are smaller than the size of the symbols.}
\label{gtcspit}
\end{figure}

To better characterize the transit signal, we obtained data with higher angular resolution. On 2019 October 10 and 17, we observed transits with three small privately-operated telescopes, revealing that the white dwarf dims by up to 56\% for eight minutes.  On 2019 October 22, we observed a transit with two larger telescopes, the Telescopio Carlos S\'anchez and Gran Telescopio Canarias (Figure \ref{gtcspit}). Together, these data show that a Jupiter-sized object transits the white dwarf in a grazing configuration (that is, the companion only occults part of the much smaller star). 

\begin{figure}[h]
\centering
\includegraphics[width=\textwidth]{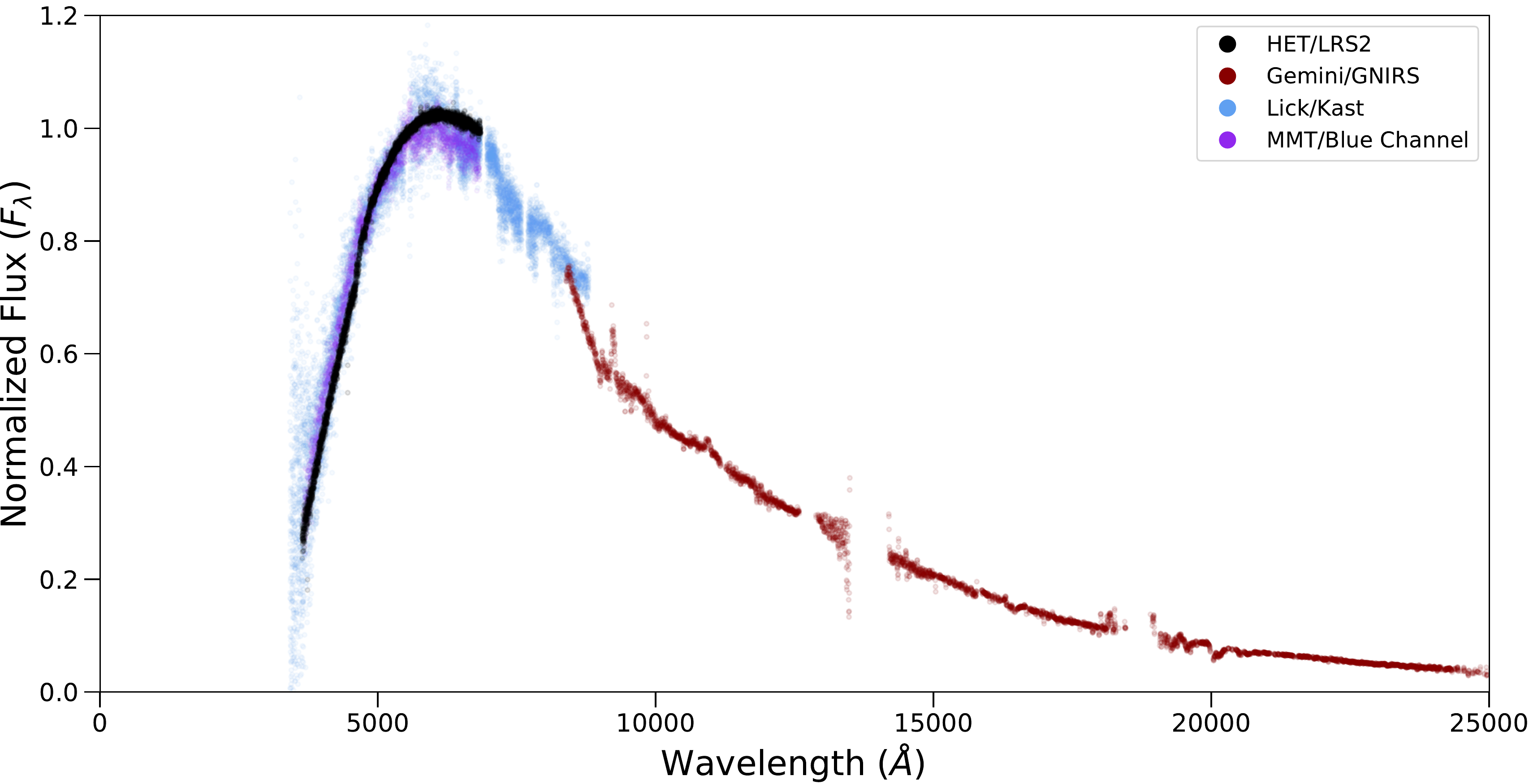}
\caption{\bedittwo{Spectroscopic observations of WD 1856. We show spectra from four observatories (the HET, the Gemini-North telescope, the Lick Shane telescope, and the MMT observatory) that have been scaled to remove offsets in their absolute flux calibrations.} The optical spectra show a pure continuum, confirming the DC spectral classification, while the near infrared spectrum from Gemini-North shows only spurious features due to imperfect correction of the telluric absorption and sky emission from Earth's atmosphere.}
\label{fig:spectra}
\end{figure}

Jupiter-sized objects can have a wide range of masses, ranging from giant planets (with masses as low as $\sim$ 0.1 \mjup) to low-mass stars ($\sim$100 \mjup). Determining the mass is usually achieved through precise Doppler monitoring of the primary star. However, the spectrum of WD 1856 is classified as type DC\cite{mccooksion}, a featureless continuum with no strong optical absorption or emission features. Optical and near-infrared spectra from the MMT Telescope, Lick Shane Telescope, Gemini-North telescope, and Hobby Eberly Telescope confirmed this classification \bedittwo{(Figure \ref{fig:spectra})}. The lack of strong spectroscopic absorption features precludes precise Doppler observations.

Instead, we constrained the mass of the transiting body based on the lack of any detectable thermal emission. We observed a transit on 2019 December 16 with NASA's \Spitzer\ Space Telescope operating at wavelengths between 4 and 5 microns. At these infrared wavelengths, the thermal emission from a low-mass star or brown dwarf would make a larger fractional contribution to the total light than at the optical wavelengths of our other observations. This, in turn, would cause the fractional loss of light during transits to be smaller at infrared wavelengths than at optical wavelengths \bedit{(absent slight differences in the stellar limb darkening profile between the two bands)}. Figure \ref{gtcspit} compares the infrared and optical light curves. There is no discernible difference in the fractional loss of light; any thermal flux from the transiting body can be no more than 6.1\% of the flux from the white dwarf (with 95\% confidence).

\begin{figure}[ht!]
\centering
\includegraphics[width=0.99\textwidth]{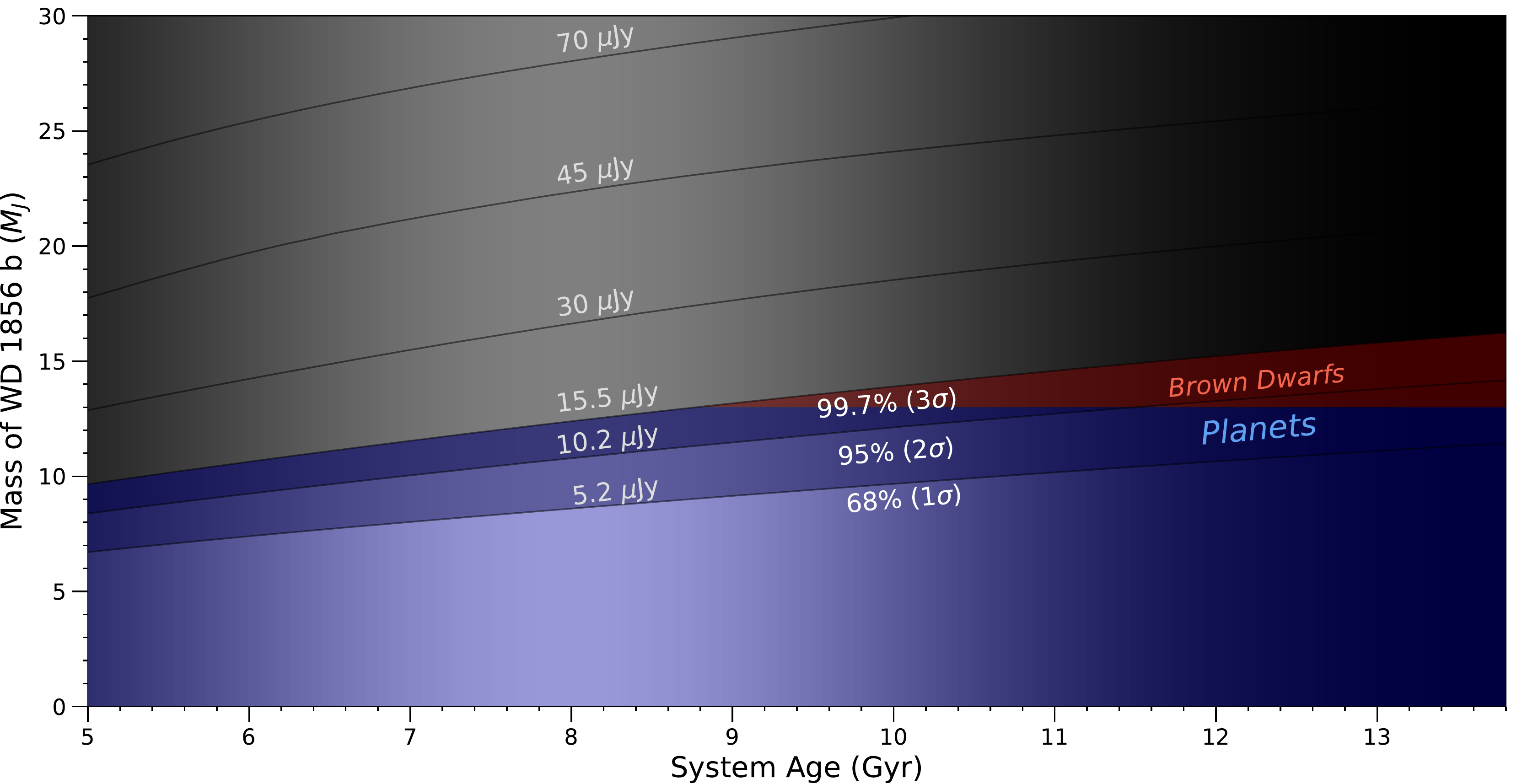}
\caption{Allowed mass range for WD 1856 b as a function of the system age. Giant planets and brown dwarfs cool and contract as they age, so higher masses are allowed by our \Spitzer\ observations for older systems. The masses and ages comprising the greyed-out region at the top of the plot (high masses) are excluded by the lack of any detectable thermal emission
with \Spitzer. The blue and red regions are the allowed ranges for planet and brown dwarf solutions, respectively, and are separated by the traditional 13 \mjup\ deuterium burning limit. The 1$\sigma$ (68\% confidence),  2$\sigma$ (95\%), and  3$\sigma$ (99.7\%) regions are shaded with darker regions representing increasingly unlikely solutions. \bedit{Several additional contours of constant brightness in the \Spitzer\ 4.5 $\mu m$ band are shown and labeled.} To convey that the system's most likely age is $\lesssim$ 10~Gyr, the background has been shaded darker for much older ages.}
\label{fig:bdmassage}
\end{figure}

Such a faint object can only be a planet or a very low-mass brown dwarf, based on theoretical models of brown dwarf evolution\cite{nelson2018} and atmospheres\cite{sonora}. Figure \ref{fig:bdmassage} shows the resulting constraints on the mass of
the transiting companion as a function of the system age. A mass exceeding 13.8 \mjup\ is ruled out regardless of age (95\% confidence), and the constraints for younger systems are even stronger. The system's motion through space suggests it is a member of the Galaxy's thin disk, implying an age less than about 10 Gyr and a mass less than 11.7  \mjup\ (95\% confidence). Therefore, the transiting body almost certainly has a mass in the planetary regime\cite{deuterium}.

Most or all of the usual circumstances that sometimes result in ``false positive'' transiting exoplanet detections can be ruled out, given the data at hand. The ground-based transit observations confirm that the \TESS\ signal is not an instrumental artefact or contamination from a different source. \bedittwo{The transit duration is too long for the companion to be another white dwarf in either a 1.4 or 2.8-day period orbit.} There is no evidence for unresolved blended sources in archival images or in the astrometric data from ESA's \Gaia\ mission. Even if there were a faint undetected companion, the transits are deep enough ($>$50\%) that they must originate from WD 1856. \bedittwo{Furthermore, the $>$50\% transit depth implies that the signal also cannot be primary and secondary eclipses of an equal-temperature white dwarf/white dwarf binary.} We conclude that WD 1856 is orbited by either a giant planet or a very low-mass brown dwarf, which we designate
WD 1856\,b.

To avoid destruction when WD 1856's progenitor evolved into a red giant, WD 1856 b must have been farther than about 1 AU from its host star, raising the question of how it arrived in the close orbit we observe today. 
Most short-period white dwarf binaries, including the handful of known white dwarf/brown dwarf pairs\cite{casewell2012, littlefair14, wd1202, parsons2017}, are believed to have formed via common envelope evolution\cite{Paczynski1976}. In this \bedit{theory}, an expanding giant star grows large enough to engulf a lower-mass binary companion. Friction from the giant star's gaseous envelope causes the companion to rapidly spiral inward towards the giant's dense core, depositing its orbital energy into the envelope. If the companion and core have enough gravitational potential energy, the envelope can be ejected, halting the companion's orbital evolution and resulting in a binary system with an orbital period ranging from hours to days. If there is not enough gravitational potential energy to unbind the envelope, then the companion continues spiraling inward towards the giant star's core until they merge.

It is difficult to explain WD 1856 b's current orbit with standard common envelope theory.  Compared to a list\cite{nelson2018} of known close white dwarf/brown dwarf binaries that were thought to have formed via common envelope \bedit{evolution}, WD 1856 b has by far the combination of lowest mass and longest orbital period of any similar system. This implies that the gravitational potential energy released during the common envelope phase is very small, which in turn makes it difficult to successfully eject the envelope of the WD progenitor.  The amount of gravitational potential energy to be released is
\begin{equation}
\Delta \phi \simeq -\frac{G M_{\rm wd} M_{\rm com}}{a} = \left(\frac{2 \pi G}{P}\right)^{2/3} \frac{M_{\rm wd}M_{\rm com}}{\left(M_{\rm wd}+M_{\rm com}\right)^{1/3}} \, \appropto \, M_{\rm com}\left(M_{\rm wd}/P\right)^{2/3}
\end{equation}
where $M_{\rm wd}$, $M_{\rm com}$, $a$, and $P$ are the WD mass, companion mass, orbital separation, and orbital period, 
respectively, after the common envelope. The brown dwarfs in the systems compiled in ref\cite{nelson2018} tend to have masses of at least 50-60 $M_J$ and orbital periods in the range of $\sim$1-4 hours. WD 1856 b's low mass ($\lesssim 14 M_J$) and long orbital period ($\sim$34 hr) could therefore have released only $\sim$15 times less gravitational potential energy than the other systems listed in ref\cite{nelson2018}.  More formally, we calculated that \bedit{throughout most of the progenitor's giant phases}, WD 1856 b's gravitational potential energy release was insufficient to eject the progenitor giant star's envelope and avoid merging with its core (see \textit{Methods}). Some groups have suggested that the envelope’s own internal energy could contribute to its ejection\cite{xu2010}, but even this extra energy source appears insufficient for WD 1856 b to have ejected the envelope. WD 1856 b can likely only have formed by this mechanism if the common envelope phase began after much of the envelope’s mass had already been lost. \bedit{Given the difficulty in forming WD 1856 b via common envelope evolution and the degree to which it stands out from the population of known post-common envelope binaries, we conclude that the system's current configuration most likely formed via some other mechanism.}

\bedit{Instead, a more likely formation history is that} WD 1856 b was a planet that underwent dynamical instability. It is well established that when stars evolve into white dwarfs, their previously stable planetary systems can undergo violent dynamical interactions\cite{debes2002, verasunpacking} that excite high orbital eccentricities. We have confirmed with our own simulations that WD 1856 b-like objects in multi-planet systems can be thrown onto orbits with very close periastron distances. If WD 1856 b were on such an orbit, the orbital energy would have rapidly dissipated due to tides raised on the planet by the white dwarf\cite{Goldreich1966, Veras2019a}. The final state of minimum energy would be a circular short-period orbit. WD 1856's advanced age ($\approx$ 5.85 Gyr) gives plenty of time for these relatively slow ($\sim$ Gyr) dynamical processes to take place. In this \bedit{case}, it is no coincidence that WD 1856 is one of the oldest white dwarfs observed by \TESS.

Future observations should be able to confirm the planetary nature of WD 1856 b or (less likely) show that it is a low-mass brown dwarf. The amplitude of features in a planet's transmission spectrum depend inversely on the strength of its surface gravity. If WD 1856 b has a mass close to that of Jupiter, its spectral features could have amplitudes of about 1\%. However, weak spectral features do not necessarily imply a large mass for WD 1856 b, because spectral features can be muted by high altitude clouds or hazes\cite{kreidberg14}. Another path to measuring WD 1856 b's mass would be to replicate our \Spitzer\ observations with the upcoming \jameswebb\ Space Telescope (\JWST). With its much larger collecting area, a single \JWST\ transit observation should either detect thermal emission from WD 1856 b or place a strong enough constraint on its mass to solidify its planetary nature.

WD 1856 b will be a focus of future observational and theoretical studies. If the object's mass is low enough for it to cool to its equilibrium temperature (about 165 K), transmission spectroscopy observations could probe species like methane and ammonia in the atmosphere of one of the coldest known transiting planets\cite{akeson2013}. 
If instead WD 1856 b has a higher mass and has retained some of its primordial heat, the white dwarf's low luminosity means infrared observations with \JWST\ could reveal WD 1856 b's thermal emission spectrum with unusual detail. Regardless of its exact mass, WD 1856 b demonstrates that low-mass objects can migrate into close orbits around white dwarfs while avoiding total tidal disruption. Unlike common envelope evolution, which predicts that low-mass objects will merge with their host star's core, there is no reason why the dynamical mechanisms we invoke to explain WD 1856 b's formation could not also be applied to even smaller planets, similar in size to Earth\cite{agol2011}.

\begin{table}
\begin{threeparttable}
\scriptsize
\centering
\caption{Summary of parameters for WD 1856+534 system.} \label{parameters}
\medskip
\begin{tabular}{lccc}
\hline
Parameter & Value & Value (Eccentric Fit) & Source \\
\hline
\textbf{Other Designations} & & &  \\
TIC 267574918& &  & \\
TOI 1690& &  & \\
LP 141-14 & &  & \\
2MASS J18573936+5330332 & &  & \\
\Gaia\ DR2 2146576589564898688 & &  & \\
\textbf{Astrometric parameters} & & &  \\
Right Ascension & 18:57:39.34&  & \Gaia\\
Declination & +53:30:33.3 &  & \Gaia\\
Right ascension proper motion & 240.759 $\pm$ 0.148 mas/yr & & \Gaia\\
Declination proper motion & -52.514 $\pm$ 0.143 mas/yr & & \Gaia\\
Parallax & 40.3983 $\pm$ 0.0705 mas & & \Gaia\\
Distance to Star & 24.754 $\pm$ 0.044 parsec & & \Gaia\\

\multicolumn{4}{l}{\textbf{Literature and New Photometric measurements}} \\
$g$ & 17.6038  $\pm$  0.0046 & & Pan-STARRS\\
$r$ & 16.9085  $\pm$  0.0025 & & Pan-STARRS\\
$i$ & 16.6248  $\pm$  0.0038 & & Pan-STARRS\\
$z$ & 16.5182  $\pm$  0.0032 & & Pan-STARRS\\
$y$ & 16.4685  $\pm$  0.0064 & & Pan-STARRS\\
$G$ & 16.9580  $\pm$  0.0010 & & \Gaia\\
$B_P$ & 17.5032 $\pm$  0.0059 & & \Gaia\\
$R_P$ & 16.2780  $\pm$  0.0033 & & \Gaia\\

$J$ & 15.677  $\pm$  0.055 & & 2MASS\\
$H$ & 15.429 $\pm$ 0.094 & & 2MASS\\
$K$ & 15.548 $\pm$ 0.186 & & 2MASS\\
$W1$ & 15.011 $\pm$ 0.027 & & ALLWISE\\
$W2$ & 15.156 $\pm$ 0.048 & & ALLWISE\\
$W3$ & $>$13.404 (2$\sigma$) & & ALLWISE\\
$W4$ & $>$9.639 (2$\sigma$) & & ALLWISE\\
\bedit{IRAC 4.5$\mu m$} & \bedit{15.042 $\pm$ 0.066} & & \bedit{this work}\\

\textbf{White Dwarf Stellar Properties} & & &  \\
Mass ($M_\star$) & 0.518$\pm$ 0.055 \msun& & this work\\
Radius ($R_\star$) & 0.0131$\pm$ 0.00054 \rsun&  & this work\\
Radius ($R_\star$) & 1.429$\pm$ 0.059 \rearth\ &  & this work\\
Surface Gravity ($\log{g_{\rm cgs}}$)& 7.915 $\pm$ 0.030& & this work\\
Effective Temperature ($T_{\rm eff}$)& 4710 $\pm$ 60 K& & this work\\
Cooling Age ($t_{\rm cool}$)& 5.85 $\pm$ 0.5 Gyr& & this work\\
Calcium abundance ($\log{\rm Ca/(H+He)}$)& $< -11.1$& & this work\\
Iron abundance ($\log{\rm Fe/(H+He)}$)& $< -8.8$& & this work\\
Magnesium abundance ($\log{\rm Mg/(H+He)}$)& $< -7.9$& & this work\\
Sodium abundance ($\log{\rm Na/(H+He)}$)& $< -10.3$& & this work\\
Sulphur abundance ($\log{\rm S/(H+He)}$)& $< -3.3$& & this work\\
\textbf{Planet Candidate Properties} & & &  \\
Orbital Period ($P$)$^{*}$ &\perplb\uperplb\ days& \perplbecc\uperplbecc\ days& this work\\
Time of Transit ($t_{t}$) &\ttransitb\uttransitb\  BJD\_TDB& \ttransitbecc\uttransitbecc\ BJD\_TDB& this work\\
Radius Ratio ($R_p/R_\star$) & \rprstb\urprstb  &  \rprstbecc\urprstbecc\ & this work\\
Scaled semimajor axis ($a/R_\star$)  &\arstb\uarstb  & \arstbecc\uarstbecc & this work\\
Semimajor axis ($a$)  &\ab\uab\ AU  & \abecc\uabecc\ AU & this work\\
Orbital inclination ($i$) &\inclb\uinclb\  deg& \inclbecc\uinclbecc\ deg& this work\\
Orbital eccentricity ($e$) & 0 &\eccb\ (2$\sigma$) & this work\\
Transit Duration ($t_{14}$) &\tdurb\utdurb\ min& \tdurbecc\utdurbecc\ min& this work\\
Planet Radius ($R_p$) &\rplb\urplb\ \rearth&\rplbecc\urplbecc\ \rearth& this work\\
Transit impact parameter ($b$) &\impb\uimpb\  & \impbecc\uimpbecc\ & this work\\
Incident Flux ($S$) &\fluxb\ufluxb\ $S_\oplus$ &\fluxbecc\ufluxbecc\ $S_\oplus$& this work\\
Equilibrium Temperature ($T_{eq}$)$^{**}$ &\teqb\uteqb\ K&\teqbecc\uteqbecc\ K& this work\\
\Spitzer\ Dilution Parameter ($d$) &\dspit\udspit\ &\dspitecc\udspitecc\ & this work\\
\bedit{Apparent IRAC 4.5$\mu m$ magnitude} & \bedit{$>$18.1 (2$\sigma$)} & \bedit{$>$18.1 (2$\sigma$)} & \bedit{this work}\\
\bedit{Absolute IRAC 4.5$\mu m$ magnitude} & \bedit{$>$16.1 (2$\sigma$)} &\bedit{$>$16.2 (2$\sigma$)} & \bedit{this work}\\

\hline
\end{tabular}
    \begin{tablenotes}
      \scriptsize
      \item \bedit{The reported uncertainties represent 68\% confidence intervals (1$\sigma$) unless stated otherwise.} 
      \item *The reported orbital period is the value measured by observers in our Solar System's barycentric frame (i.e. slightly Doppler shifted from the orbital period in the WD 1856 system's rest frame).
      \item **Equilibrium temperature $T_{eq}$ calculated assuming an albedo $\alpha$ uniformly distributed between 0 and 0.7 and perfect heat redistribution. $T_{eq}=T_{\rm eff}(1 - \alpha)^{1/4}\sqrt{\frac{R_\star}{2a}}$.
    \end{tablenotes}
  \end{threeparttable}
\end{table}

%Photoevaporation?
\newpage
\begin{methods}

\renewcommand{\figurename}{Extended Data Figure}
\renewcommand{\tablename}{Extended Data Table}
\setcounter{table}{0}  
\setcounter{figure}{0}  

\subsection{\TESS\ Target Selection and Observations:}
We discovered the transits of WD 1856 b in data from NASA's \TESS\ mission\cite{ricker2015}. \TESS\ is a satellite which observes a 96$^\circ$ by 24$^\circ$ region of sky with four 10 cm optical cameras. \TESS\ observes the same region of sky continuously for approximately 28 days at a time; each 28 day observation is called a sector. Over the course of its two-year prime mission, \TESS\ will observe 26 sectors, covering over 70\% of the sky. \TESS\ collects and downloads images of its entire field of view with 30-minute exposure times, but \TESS\ also observes 20,000 carefully chosen targets each month with shorter (two minute) exposure times. Because transits of white dwarf stars typically have durations much shorter than the 30-minute cadence of \TESS's full frame image downloads, we proposed for two-minute-cadence observations of known and candidate white dwarf stars.

We first proposed \TESS\ observations of white dwarf stars in the Southern ecliptic hemisphere in late 2017, before the second data release (DR2) from ESA's \Gaia\ mission  enabled the discovery of hundreds of thousands of new white dwarf candidates. We proposed two-minute cadence observations of white dwarfs in the Montreal White Dwarf Database (MWDD)\cite{dufour} brighter than a magnitude of 17.5 in either $V$, $I$, or \TESS\ bands and which are more than 20$''$ from any brighter stars which would contaminate the TESS photometric apertures. We also performed our own search (using the same $V$ or $I$ or \TESS\ $\leq$ 17.5 magnitude limit) for and proposed observations of new candidate white dwarfs by finding hot stars with high Reduced Proper Motion (RPM) - a proxy for luminosity\cite{gould2003}. We used proper motions from the Hot Stuff for One Year catalog\cite{altmann}, \Gaia\ $G$-band magnitudes, and 2MASS $J$-band magnitudes to calculate each star’s RPM. We defined cuts in color/RPM space to select likely white dwarfs. A total of 615 unique white dwarf candidates from our program were observed during \TESS's first year of operations. 

For the second year of \TESS\ observations of the Northern ecliptic hemisphere, we identified targets from a catalog of candidate white dwarfs\cite{GentileFusillo} based on \Gaia\ DR2. We proposed two-minute observations of all white dwarf candidates brighter than Gaia $G$-band magnitude of 17 with a greater than 75\% probability of being a true white dwarf, and removed white dwarfs less than 20$''$ from any brighter stars which would contaminate the TESS photometric apertures. Thanks to \Gaia\ DR2, our Northern target list was much more complete than our Southern list. So far (through Sector 19), a total of 1189 unique Northern white dwarf candidates from our program have been observed.  

Once the \TESS\ data on these targets were collected and downlinked from the spacecraft, they were processed by the Science Processing Operations Center (SPOC) pipeline\cite{Jenkins:2015, Jenkins:2016} based at NASA Ames Research Center. The SPOC pipeline performed pixel-level calibrations, identified optimal photometric apertures, extracted light curves, corrected for systematic errors and diluting flux from nearby stars \cite{smith, stumpe}, and searched for periodic transit signals\cite{jenkins:2002}. The SPOC pipeline's periodic transit search algorithm detected a convincing, 1.4 day period, short-duration transit signal around WD 1856 (listed in the \TESS\ Input Catalog as TIC 267574918). The transits were first detected in \TESS's Sector 14 observations, but the signal was rejected by an automatic classification algorithm designed to separate viable planet candidates from false positives\cite{guerrero}. We noticed WD 1856 in a visual inspection of all possible transit-like signals around white dwarfs identified by the SPOC pipeline (including those rejected by the automatic classifier), and initiated follow-up observations. Subsequently, WD 1856 was also observed in \TESS\ Sector 15 and Sector 19 (and will be observed again in Sector 22 and 26). The transits were re-detected in a combined analysis of the Sector 14-15 data and in the Sector 19 data. After being rejected by the automatic classifier in Sectors 14 and 15, WD 1856 b's transit signal was promoted to the status of ``planet candidate'' in the Sector 19 observations and was given the designation \TESS\ Object of Interest (TOI) 1690.01. 

Though the \TESS\ data confidently revealed the presence of 6-8 minute long, 1.4 day period transits, and tests performed by the SPOC pipeline showed that the signal likely originated on WD 1856 (and not on some other nearby star), the \TESS\ light curve data were challenging to interpret. Compared to many other ground-based or space-based telescopes, \TESS\ has relatively poor spatial resolution. \TESS's optics focus about 50\% of a given star's light into one of its 20\arcsecond\ pixels, and the wings of the point spread function (PSF) extend several pixels farther. This poses challenges for observations of faint stars like WD 1856, especially since it is only about 40\arcsecond\ (2 pixels) away from a pair of physically associated M-dwarf stars (see below). The M-dwarfs are about 100 times brigher than WD 1856 in the \TESS\ bandpass and contribute a significant amount of flux into WD 1856's photometric aperture. In situations like this, the dilution correction applied by the SPOC pipeline to the WD 1856 light curve is fairly uncertain given the difficulty in precisely measuring the wings of the \TESS\ PSF. This uncertainty in the SPOC dilution correction translated to a substantial uncertainty in the true depth of WD 1856 b's transits. 

WD 1856 stands out among the stars targeted in our \TESS\ sample as one of the coolest, and therefore oldest, white dwarfs we observed. Among the 1724 white dwarfs in our sample observed by \TESS\ in Sectors 1-19 with catalog reported effective temperatures\cite{GentileFusillo}, only 8 white dwarfs are cooler than WD 1856.

\subsection{Archival Imaging and Search for Companions}
We searched for both wide and close stellar companions to WD 1856 in archival survey data. WD 1856 was previously believed\cite{mccooksion} to be part of a visual triple star system with a pair of M-dwarfs called  G 229-20. G 229-20 consists of two nearly equal-brightness M-dwarf stars separated by about 2.3 arcseconds ($\approx 56$ AU projected separation). The M-dwarf pair is located approximately 43\arcsecond\ away from WD 1856 ($\approx$ 1000 AU projected separation). Data from \Gaia\ DR2 show that G 229-20 A/B have nearly identical proper motions and parallaxes to WD 1856, confirming the three stars are physically associated. From here on, we refer to the Northern component of the binary as G 229-20 A since it is slightly brighter in resolved photometry from \Gaia\ DR2.

We searched for additional co-moving companions in the \Gaia\ archive. We queried all stars in \Gaia\ DR2 within 600\arcsecond\ of WD 1856 (approximately 15000 AU projected separations) and looked for proper motions similar to WD 1856 and G 229-20 A/B. We found no stars with remotely similar space motions to the WD 1856 system.

We also checked to see if the \Gaia\ observations showed any evidence for close, unresolved companions to either WD 1856 or G 229-20 A/B. Sometimes, close binary companions can introduce excess scatter into the \Gaia\ astrometric observations\cite{evans, rizzuto}. This excess scatter is parameterized in a statistic called the Renormalized Unit Weight Error (RUWE\cite{ruwe}). Solutions with low astrometric scatter have RUWE values close to 1, while stars whose astrometric solutions show anomalously high scatter (perhaps due to astrometric motion from an unresolved binary companion) tend to have RUWE values greater than about 1.4. None of the members of the WD 1856 system show evidence for excess astrometric scatter that might reveal close companions; the RUWE values for WD 1856,  G 229-20 A, and  G 229-20 B are 1.04, 1.01, and 0.94 respectively. 

Finally, we searched for background stars at the present-day position of WD 1856 in archival imaging. WD 1856 was observed in the Palomar Observatory Sky Survey (POSS) on 27 July 1952 with a photographic plate with a blue sensitive emulsion. Due to its high proper motion, WD 1856 has moved over 16 arcseconds since being imaged by POSS, making it possible to search for background stars at WD 1856's present-day position. There are no possible background contaminants at WD 1856's current position brighter than the POSS image's limiting magnitude (approximately 21$^{\rm st}$ magnitude in blue\cite{abell1955}). Extended Data Figure \ref{apertures} shows the POSS image of WD 1856 along with modern images from Pan-STARRS and \TESS. 

\begin{figure}                                             
\includegraphics[width=1.0\textwidth]{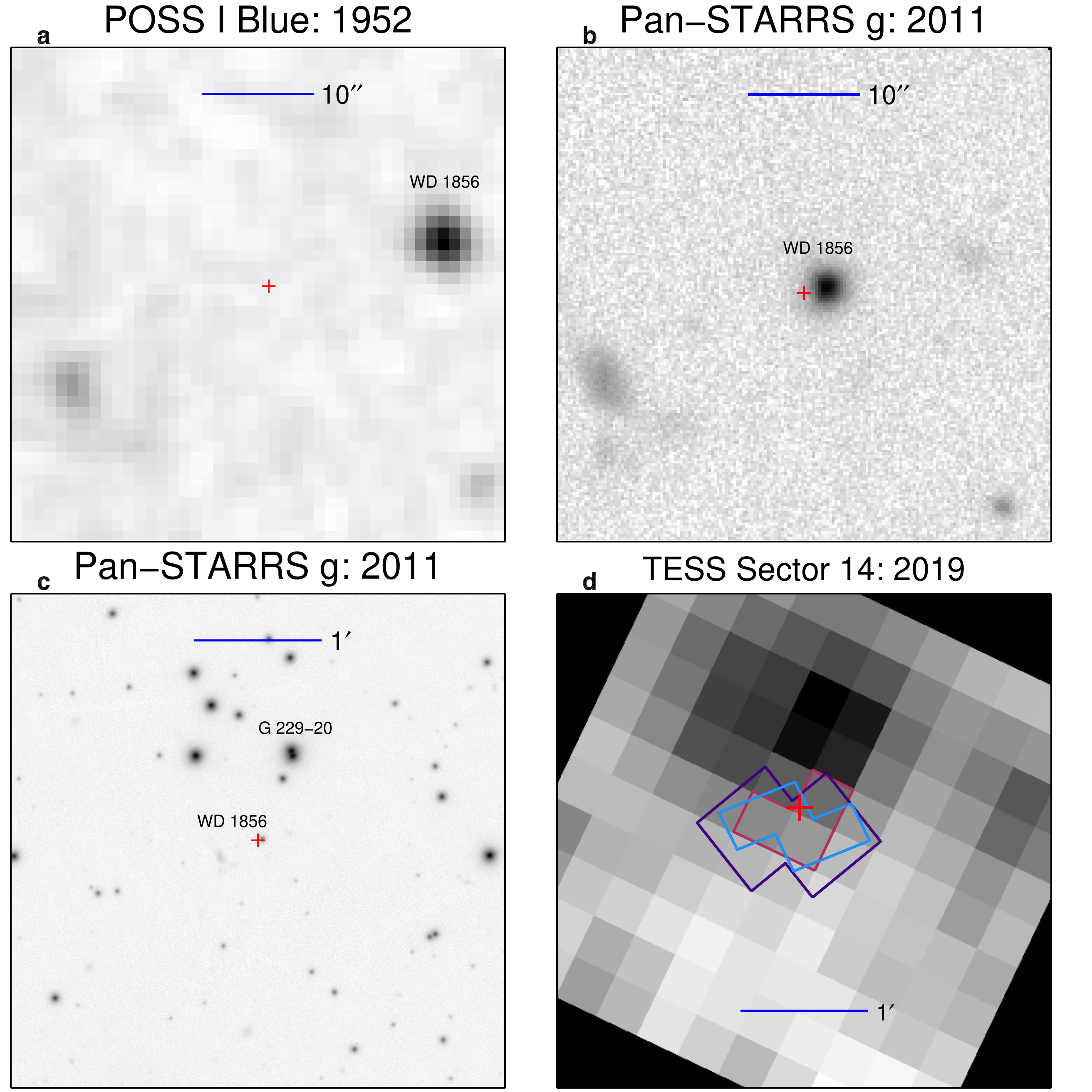} 
\caption{Archival imaging of WD 1856. \textbf{a}, From the Palomar Observatory Sky Survey on a photographic plate with a blue-sensitive emulsion. \textbf{b}, From the Panoramic Survey Telescope and Rapid Response System (Pan-STARRS) survey in $g$ band. \textbf{c}, From the Pan-STARRS survey in $g$ band, zoomed out to show the co-moving M-dwarf pair (labeled G 229-20). \textbf{d}, Co-added \TESS\ image from Sector 14. The photometric apertures for the three sectors of TESS observations (14, 15, and 19) are shown as red, purple, and blue colored outlines, respectively. The present-day location of WD 1856 is shown with a red cross in all images.  }
\label{apertures}                                              
\end{figure} 

\subsection{Ground-based Transit Follow-up}
Based on the orbital period and time of transit inferred from the \TESS\ observations of WD 1856, we planned ground-based transit observations to confirm the transit signal and measure its true depth. We observed transits of WD 1856 b on 10 October 2019 and 17 October 2019 (UTC) with three small privately owned ground-based telescopes in Arizona: a 16-inch telescope at the Hereford Arizona Observatory (operated by Bruce Gary), a 16 inch telescope at Raemor Vista Observatory, and a 32-inch at Junk Bond Observatory (both operated by Thomas G. Kaye). We observed in white optical light without any color filter; our effective bandpass was defined by the telescope systems' throughput and the CCDs' quantum efficiency. Weather conditions on both nights were clear and stable. The data were reduced following standard procedures for these telescopes\cite{rappaport2016}. All three telescopes confidently detected the transit signal with a consistent $\approx$60\% depth on both nights.\bedittwo{ The data showed that the depths of odd and even numbered transits are indistinguishable and both greater than 50\% of the total brightness, so WD 1856 must not be a nearly equal-brightness eclipsing binary star with a true orbital period of 2.8 days (since the sum of the depths of a binary's primary and secondary eclipse cannot exceed 100\%).}

After confirming the transits and determining the depth, we observed another transit of WD 1856 b with two larger telescopes to more precisely determine the transit shape and attempt to detect or rule out any color dependence in the transit depth. We observed a transit of WD 1856 on 22 October 2019 with the MuSCAT2 instrument\cite{muscat2} on the 1.52 meter Telescopio Carlos S\'anchez and with the Optical System for Imaging and low-Intermediate-Resolution Integrated Spectroscopy (OSIRIS) imager/spectrograph on the 10.4 meter Gran Telescopio Canarias (GTC). MuSCAT2 provides simultaneous multi-color images of a 7\farcm4$\times$7\farcm4 field of view with fast readout times. We observed in four bands simultaneously: $g$, $r$, $i$, and $z_s$. We reduced the observations with the standard MuSCAT2 pipeline and detected the transit with the same depth in each of the four MuSCAT2 bands. Our GTC observations used OSIRIS as an imager to obtain a precise $g^\prime$-band light curve of WD 1856. We obtained 10-second exposures of WD 1856 and read out the detector in frame transfer mode, which allowed us to observe nearly continuously (one frame was being read out while the next was exposing). We reduced the observations using standard IRAF scripts to calibrate the images and extract light curves for both WD 1856 and comparison stars. We experimented with different sized photometric apertures, and found that a 6 pixel aperture minimized the scatter in the light curve. The resulting light curve was extremely precise (0.5\% scatter per 10 second exposure) and revealed a smooth, symmetric 56\% deep transit. 

Our follow-up light curves are shown in Extended Data Figure \ref{allground}, compared to the \TESS\ discovery light curve (corrected for the dilution from nearby stars).

\begin{figure}[t!]
%\epsscale{.6}
  \begin{center}
      \leavevmode
              \begin{minipage}[c]{0.6\textwidth}

\includegraphics[width=1.0\textwidth]{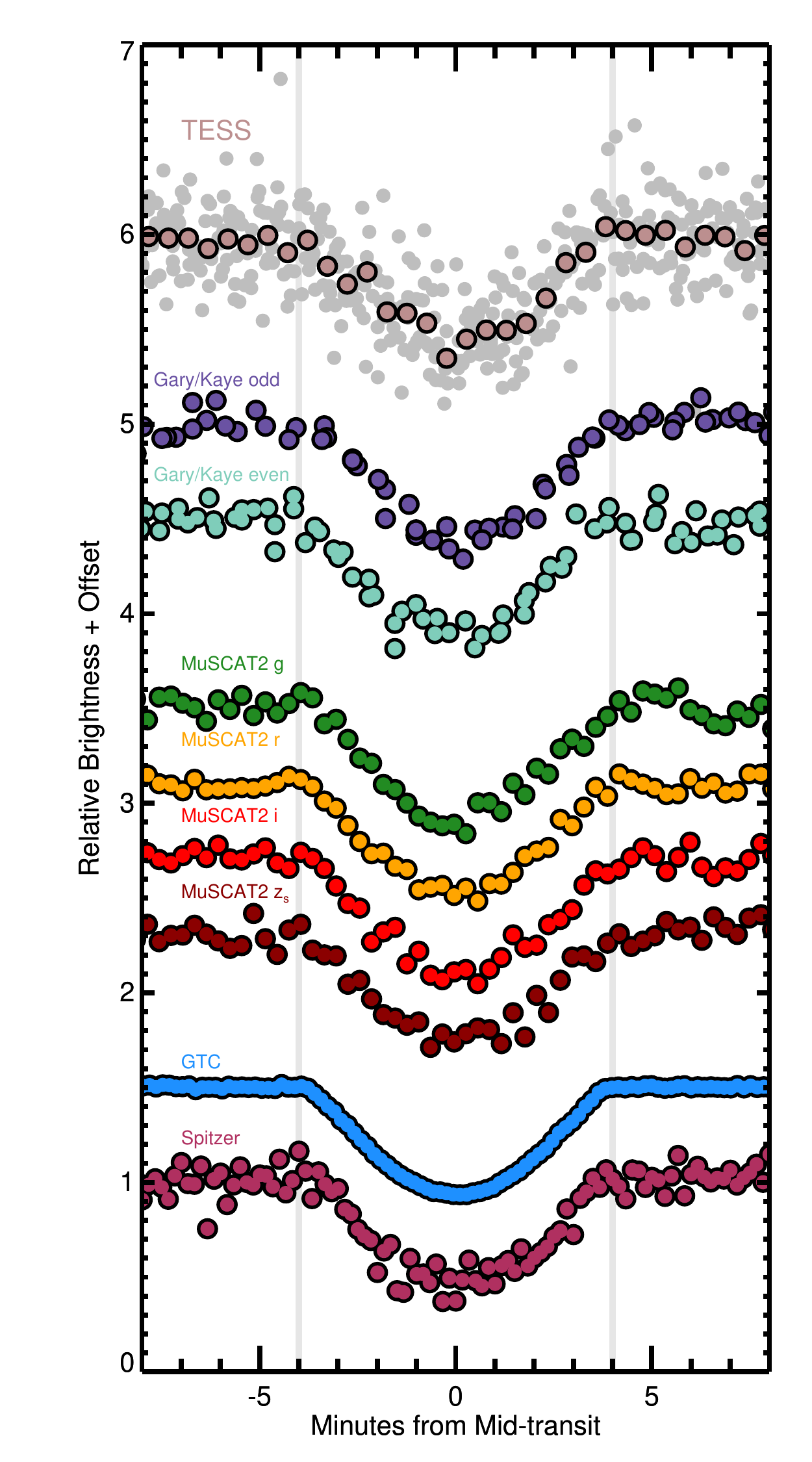}
  \end{minipage}\hfill
  \begin{minipage}[c]{0.39\textwidth}

\caption{\bedittwo{All transit observations of WD 1856}. From top to bottom, we show the light curves (arbitrarily offset for visual clarify) from \TESS; data from several private telescopes in Arizona (operated by Gary and Kaye) with odd and even-numbered transits shown separately; simultaneous light curves in four colors from MuSCAT2; a light curve from the GTC, and a light curve from \Spitzer. The individual two-minute-cadence \TESS\ flux measurements are shown as grey points, and the rose-colored points are averages of the brightness in roughly 30 seconds in orbital phase. The \TESS\ data have been corrected for dilution from nearby stars so that the transit depth matches that of the GTC data. 
}\label{allground}
  \end{minipage}

\end{center}
\end{figure}

\subsection{Spectroscopy of WD 1856}
A previous study assigned WD 1856 the spectral type classification of DC, indicating a continuum dominated spectrum with very few weak absorption features\cite{mccooksion}. We sought to confirm this classification and detect any weak absorption features by collecting our own optical spectroscopic observations.  We observed WD 1856 on 5 October 2019 with the Blue Channel spectrograph\cite{schmidt} on the 6 meter MMT telescope at Fred L. Whipple Observatory. We used the 500 line/mm grating and achieved 3.8 \AA\ spectral resolution over a bandpass from 3700-6800 \AA. A 10 minute exposure yielded a signal-to-noise ratio of about 50 per pixel or 80 per resolution element. The resulting spectrum confirmed the DC spectral classification.

We continued searching for features in WD 1856's spectrum by extending our wavelength coverage beyond the red limits of our MMT Blue Channel observations. We obtained 60 minute exposures of WD 1856 on both 11 October 2019 and 12 October 2019 with the Kast Double Spectrograph\cite{kast} on the 3 meter Shane Telescope at Lick Observatory. On both nights, we configured the blue arm of the spectrograph to yield spectra with a resolving power $R = \lambda/\Delta\lambda =$ 1300 over the wavelength range 3420-5480\AA. We changed the configuration of the red arm between the two observations; on 11 October, we observed over a bandpass from 5570 to 7860 \AA, while our 12 October observations pushed farther red from 6400 to 8800 \AA\ (both with R=3500). 

We observed WD 1856 on 30 October 2019 and 1 November 2019 with the Low Resolution Spectrograph 2 (LRS2\cite{chonis}) on the 10 meter Hobby Eberly Telescope at McDonald Observatory. LRS2 is a combination of two integral field dual-channel spectrographs: one operating in the blue (3700 to 7000 \AA) and one operating in the red (6500 to 10500 \AA). We observed WD 1856 with the two blue channels of LRS2 with a spectral resolving power of $R = \lambda/\Delta\lambda =$ 1910 from 3700-4700 \AA\ and $R=$1140 from 4700 \AA\ to 7000 \AA.  Each observation was 30 minutes in duration. The spectra were initially reduced with the automatic HET pipeline Panacea\cite{panaceawebpage}.  The pipeline performs basic CCD reduction tasks, wavelength calibration, fiber extraction, sky subtraction, and flux calibration.  We used the flux-calibrated, fiber extracted spectra for the UV and orange channels to construct a single data cube correcting for differential atmospheric refraction and the small 0\farcs3 offset between the two channels. We collapsed the datacube along the wavelength axis into an image of the LRS2 field of view, identified all fibers with at least 33\% the flux of the brightest fiber, and summed the flux in those particular fibers at each wavelength in the datacube to extract a spectrum. The LRS2 spectra had the highest signal-to-noise ratio of all of our observations, but still showed no compelling evidence for any spectral features. In particular, the LRS2 spectra rule out any \halpha\ absorption feature deeper than about 1\%. 

Finally, we observed WD 1856 on 21 November 2019 with the Gemini Near InfraRed Spectrograph (GNIRS\cite{{elias2006a,altmann}}) on the 8.1m Gemini-North telescope (program ID GN-2019B-DD-107) at Maunakea Observatory in Hawaii. The 32 l/mm grating was used in the cross-dispersed mode, which provides continuous wavelength coverage from 1.0-2.5~micron. A slit width of 1\farcs0 yielded a spectral resolving power of $R\approx$ 500. Our total exposure time was 48 minutes, broken into 12 individual exposures (three sets of four exposures offset in an ABBA pattern).  A telluric standard (HIP~95656) was observed immediately after the science observations. The observing conditions were excellent: sky was clear and seeing was $\sim$ 0\farcs35 in H band around the target. Data reduction was performed using the XDGNIRS pipeline\cite{Mason2015} v2.2.6. The correction for sky emission features and absorption due to Earth's atmosphere was imperfect and introduced some artefacts into the data, but we saw no evidence that any of the features in the data are actually spectral lines from WD 1856's atmosphere. 

Our spectra of WD 1856 are shown in Figure \ref{fig:spectra}.

\subsection{Spectroscopy of G 229-20 A/B}
We also obtained ground-based optical spectra of G 229-20 A/B, the co-moving pair of companions to WD 1856. We observed G 229-20 A and B with the Kast Double Spectrograph on the 3 meter Shane Telescope at Lick Observatory. These observations were conducted on 11 October 2019, the same night as the first of our two Kast observations of WD 1856, and were taken with an identical instrument setting (R = 1300 from 3420-5480 \AA\ and R = 3500 in the red from 5570 to 7860 \AA). Seeing conditions were good enough to resolve the two stars, so we observed them simultaneously by rotating the spectrograph slit to the position angle of the binary and placing both stars on the slit.  We extracted spectra of the two stars using standard IRAF routines. While the stars were resolved, there was still some blending along the spatial axis. 

We obtained medium-resolution spectra of G 229-20 A/B with two different echelle spectrographs. One spectrum came from the FIbre-fed Echelle Spectrograph (FIES\cite{fies}) on the Nordic Optical Telescope (NOT) on the island of La Palma, Spain on 2020 February 18. We used FIES in high-efficiency mode, in which the spectrogaph is fed with a 2\farcs5 octagonal fiber to achieve a resolving power of R=25,000. We reduced the spectra using the FIEStool pipeline\cite{fiestool}. We obtained the second spectrum with the Tillinghast Reflector Echelle Spectrograph (TRES\cite{gaborthesis}) on the 1.5 meter telescope on Mt. Hopkins, Arizona on 2020 February 24. We used the standard instrumental setup with the spectrograph fed by a 2\farcs3 fiber to achieve a spectral resolving power of R=44,000. We reduced the spectra following standard practice for this instrument\cite{buchhave2012}. We cross-correlated the spectra with an archival observation of Barnard's Star and found that the absolute radial velocity of G 229-20 A/B is 17.9 $\pm$ 0.1 \kms (on the IAU standard system\cite{stefanik}). We also inspected the \halpha\ line for G 229-20 A/B  from the FIES spectrum. G 229-20 A/B have \halpha\ in absorption, with an equivalent width of -0.32 \AA\ (where equivalent width is defined to be positive for emission features). 

We also used an archival spectrum of G 229-20 A published in a previous work\cite{lepine}. The observation was made on 25 August 2006 with the MkIII spectrograph on the McGraw-Hill 1.3 meter telescope at MDM Observatory and covered the wavelength range of 6200–8700 \AA.  The authors assigned the star a spectral type of M3.5.

\subsection{\Spitzer\ Observations}
We observed a transit of WD 1856 b with the InfraRed Array Camera (IRAC) on NASA's \Spitzer\ Space Telescope on 2019 December 16. We observed in IRAC Channel 2, the reddest possible channel (sensitive to wavelengths of light between 4 and 5 microns) to best constrain the thermal flux from a faint, cool companion. We followed standard procedures for precise photometric observations with IRAC. We began with a 30-minute long ``burn-in'' period where we obtained dithered images of WD 1856 to allow both the spacecraft and detector to settle into equilibrium prior to the actual transit observations. We then observed WD 1856 for approximately two hours surrounding the predicted time of transit from our ground-based observations. These observations were conducted in ``peak-up'' mode, where WD 1856 was carefully placed on a well-characterized pixel known to have minimal sensitivity variations. Images from a 32$\times$32 pixel subarray were collected and saved every two seconds. Finally, after the transit observation was complete, we concluded our observations with 15 minutes of dithered imaging observations of WD 1856 for calibration purposes. 

We analyzed the \Spitzer\ data with the Photometry for Orbits, Eclipses, and Transits (POET) pipeline\cite{cubrillos}. POET extracts raw light curves from the images and optimizes a transit model while simultaneously modeling and removing spacecraft systematic errors. We investigated different sizes for the photometric aperture and found the best results with a relatively small 1 pixel radius (as expected for a star as faint as WD 1856). We optimized the transit and systematics model using Markov Chain Monte Carlo (MCMC). The transit of WD 1856 was clearly detected in the \Spitzer\ observations with nearly identical characteristics to the optical transit observations. 

We also used the out-of-transit \Spitzer\ observations to measure the combined flux of WD 1856 and WD 1856 b in IRAC Channel 2.  We measured the flux using standard aperture photometry as done in previous \Spitzer\ observations of white dwarfs\cite{xu2011, xu2018} using a 2 pixel (1.2 arcsec) aperture (while applying a correction for any flux lost outside the aperture). We determined the total combined flux from WD 1856 and WD 1856 b in IRAC band 2 to be 173 $\pm$ 10 \microjanskys. We also searched for other faint red companions in the \Spitzer\ observations. We coadded the individual \Spitzer\ subarray observations to yield a deep $39^{\prime\prime}\times39^{\prime\prime}$ image of the region surrounding WD 1856 b.  We detected one faint source (at RA=18:57:39.9, Dec= +53:30:48.9), with a measured flux of 27 $\pm$ 5 \microjanskys\ without an optical counterpart. Given its distance from WD 1856 (16$^{\prime\prime}$ or 400 AU projected separation) and the M-dwarf companions (30$^{\prime\prime}$ or 750 AU projected separation), we believe the source is more likely to be a background star or galaxy than a bound companion (since the probability of a chance alignment is high). Otherwise, we find no additional sources near WD 1856 with flux greater than 16 \microjanskys\ (3$\sigma$ confidence), which at the distance of the WD 1856 system corresponds to brown dwarfs with mass $m> 16$\mjup\ (for ages up to 13.8 Gyr).

\subsection{White Dwarf Stellar Properties}

We determined fundamental stellar parameters for WD 1856 using archival photometric observations and our high signal-to-noise optical spectra from the HET. We followed the procedure of Blouin et al. (2019)\cite{blouin} and fit cool white dwarf spectral and evolutionary models\cite{blouin2018} to broad-band photometry from the Pan-STARRS and 2MASS surveys and the trigonometric parallax from \Gaia\ DR2. We modeled WD 1856's SED/spectra with atmospheres with a variety of different compositions ranging between ${\rm H/He}=10^{-5}$ and ${\rm H/He}=10^2$. We compared the predicted depth of the \halpha\ absorption feature from the different models with the observed HET spectrum (Extended Data Figure \ref{fig:halpha}); pure helium and most hydrogen/helium mixtures are consistent with our observed spectrum, but if WD 1856 b had a pure hydrogen atmosphere (or nearly so), we likely would have seen an \halpha\ absorption feature in our HET spectra.  The models with at least some helium also were a better match to the observed SED; a pure hydrogen model over-predicts WD 1856's NIR flux, while models with at least some helium better match the observations (see Extended Data Figure \ref{fig:sed}). 

We derived the white dwarf's fundamental stellar parameters from the results of our fits to the model atmospheres with varying ratios of hydrogen and helium. We found that a model with equal quantities of hydrogen and helium (50\%/50\% H/He) gave the best fit to the data. The resulting stellar parameters for some of the models we evaluated are given in Extended Data Table \ref{modelcomparison}. The fits to pure hydrogen and 50\%/50\% H/He mixture yielded fairly consistent stellar parameters, while the pure helium atmosphere gave a significantly larger white dwarf and lower stellar mass. This discrepancy is due to the effects of He-He-He collision-induced absorption (CIA) in a pure helium atmosphere, which absorbs a significant fraction of a white dwarf's infrared flux\cite{Kowalski}. However, the efficiency of this opacity source is fairly uncertain, and it is plausible that its effects are overestimated in the pure He model.

We adopt the stellar parameters from the 50\%/50\% H/He model that best matched our observations and use them throughout the rest of the paper. However, since WD 1856's atmospheric composition is not well constrained, we adopted conservative uncertainties on our stellar parameters. We inflated the formal uncertainties on the mass and radius from our model fits by adding a 10\% and 3.3\% uncertainty in quadrature, respectively. Our final, adopted values for the star's mass and radius are: $M_\star=$0.518$\pm$ 0.055 \msun\ and $R_\star=$ 0.0131$\pm$ 0.00054\rsun.

\begin{table}
\scriptsize
\centering
\caption{Comparison of White Dwarf Parameters from Different Atmosphere Models.} \label{modelcomparison}
\medskip
\begin{tabular}{lccc}
\hline
Parameter & 100\% H & 100\% He & 50\%/50\% H/He\\
\hline

Mass ($M_\star$)                        & 0.537 $\pm$ 0.018 \msun   & 0.396 $\pm$ 0.018 \msun    & 0.518 $\pm$ 0.018\\
Radius ($R_\star$)                      & 0.0131 $\pm$ 0.0014 \rsun & 0.01489 $\pm$ 0.0003 \rsun & 0.0131 $\pm$ 0.0003\\
Surface Gravity ($\log{g_{\rm cgs}}$)   & 7.931  $\pm$ 0.030        &  7.686 $\pm$ 0.030         & 7.915 $\pm$ 0.030\\
Effective Temperature ($T_{\rm eff}$)   & 4785 $\pm$ 60 K           & 4430 $\pm$ 60 K            & 4710 $\pm$ 60\\
Cooling Age ($t_{\rm cool}$)            & 5.7 $\pm$ 0.5 Gyr                   &4.25 $\pm$ 0.5  Gyr                    & 5.85  $\pm$ 0.5 Gyr\\

\hline
\end{tabular}
\end{table}

We tested how much our results depend on the specific white dwarf models used by rederiving WD 1856's stellar parameters using alternate methods. We fit\cite{stassun2018} WD 1856's spectral energy distribution (SED) with a simple blackbody curve and found a best-fit temperature of  $T_{\rm{eff}}$ = 4720 $\pm$ 50 Kelvin, a bolometric flux $F_{\rm bol}$ = 3.93$\times 10^{-12}$ $\pm$ 0.23$\times 10^{-12}$ erg\,s$^{-1}$\,cm$^{-2}$, and a stellar radius of $R_\star$ = 0.01298 $\pm$ 0.00013 \rsun. Using an approximate fitting formula\cite{eggletonbook} designed to mimic the mass/radius relation from simple zero-temperature (black dwarf) models\cite{zapolsky} and assuming a 2:1 oxygen/carbon ratio, we calculated a mass of $M_\star$ = 0.54 $\pm$ 0.01 \msun. We also estimated WD 1856's cooling age using analytic relations\cite{mestel} and found $t_{\rm cool} \sim$ 4 Gyr, with uncertainties of roughly a factor of two\cite{vanhorn}. All of these values are in good agreement with our adopted values, indicating that our results are fairly robust to different model assumptions.

Finally, we used the non-detection of spectroscopic features to place upper limits on the abundance of other elements in WD 1856's atmosphere. With our HET spectrum, we place strong limits on the presence of Ca, Fe, Mg, Na. When found in the atmospheres of white dwarfs, these elements are usually attributed to accretion from tidally disrupted rocky bodies like asteroids or small planets. Since WD 1856 b is roughly the size of Jupiter, we also searched for elements more consistent with the composition of a giant planet's atmosphere, like those recently found in the atmosphere of WD J0914+1914\cite{gansicke2019}. It is harder to constrain the abundances of these elements because they show few spectral features at wavelengths covered by our spectroscopy. We can rule out sulphur abundances greater than $\log{(S/H)}= -3.3$, but this limit is weaker than the measured sulphur abundance on WD J0914+1914. Future observations with higher spectral resolution and signal-to-noise will test whether WD 1856 shows evidence of accretion from its companion. 

\begin{figure}[h]
\centering
\includegraphics[width=\textwidth]{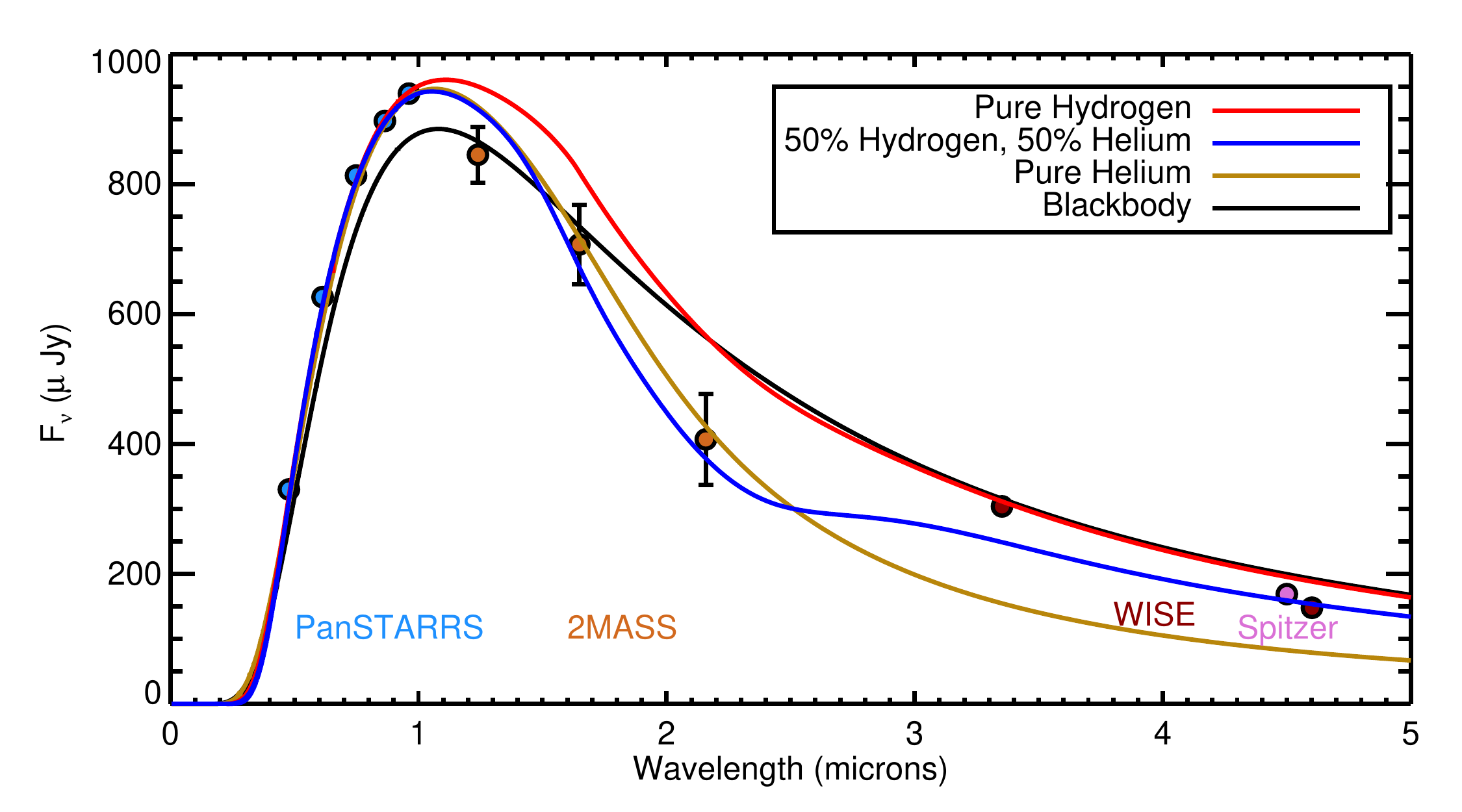}
\caption{Spectral energy distribution of WD 1856. Photometric measurements from Pan-STARRS\cite{chambers}, 2MASS\cite{twomass}, WISE\cite{allwise}, and \Spitzer, are shown as blue, orange, dark red, and pink points. The formal 1$\sigma$ \bedittwo{(standard deviation)} photometric uncertaitines on the Pan-STARRS and WISE points are smaller than the symbol size. Three different SED models are shown as solid curves: a pure hydrogen atmosphere model (red), a 50\% hydrogen, 50\% helium model (blue), and a blackbody curve (black). None of the three SED models capture all of the SED's features, but all three yield relatively consistent effective temperatures and stellar parameters.}
\label{fig:sed}
\end{figure} 

\begin{figure}[h]
\centering
\includegraphics[width=\textwidth]{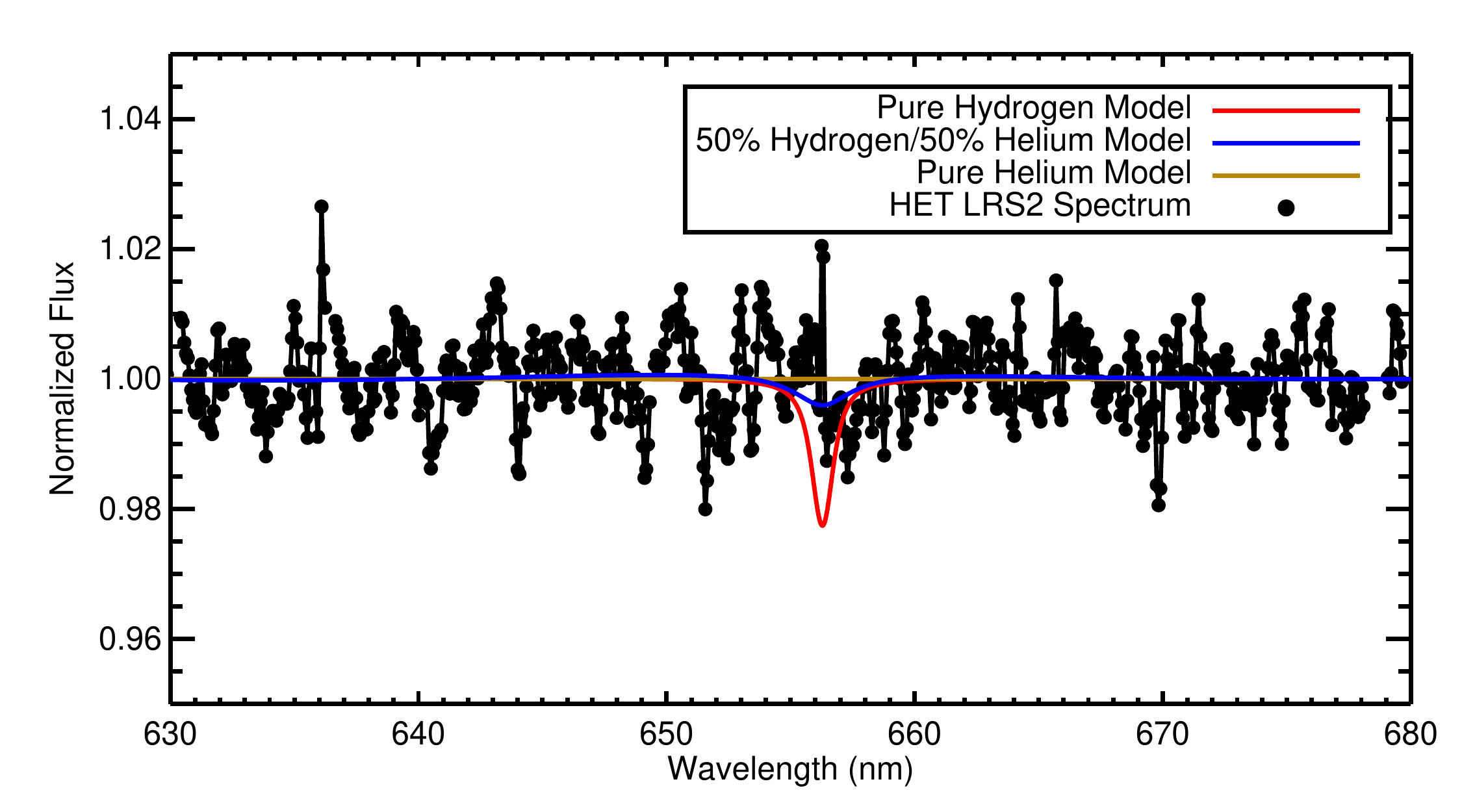}
\caption{Spectrum of WD 1856 near the \halpha\ line. Our summed HET/LRS2 spectrum (black connected points) is shown in comparison with three atmosphere models: a pure hydrogen model (red), a 50\% hydrogen, 50\% helium model (blue), and a pure helium model (gold). We likely rule out a pure hydrogen atmosphere based on our non-detection of an \halpha\ feature in our LRS2 spectra, but otherwise remain uncertain about the precise composition of WD 1856's envelope.}
\label{fig:halpha}
\end{figure} 

\subsection{M-dwarf Stellar Properties}
We determined the masses of G 229-20 A/B using broadband photometry and their \Gaia\ DR2 trigonometric parallax measurements. In most photometric surveys (including 2MASS and Pan-STARRS), G 229-20 A and B are not well resolved and only have combined flux measurements. The two stars are, however, resolved in \Gaia\ DR2 and have individually reported flux measurements. We converted the flux ratio of A/B from \Gaia\ DR2 to a flux ratio in 2MASS $K$-band using spectrophotometric standards from Mann et al. (2015\cite{mann2015}). We then estimated the mass of each star using the $M_{K_S}-M_*$ relation from Mann et al. (2019\cite{mann2019}), forcing the total $K_S$-band flux to match the unresolved measurement. This yielded masses of 0$.313\pm0.011M_\odot$ and $0.306\pm0.010M_\odot$ for A and B, respectively. The unresolved 2MASS $K_S$ measurement has a photometric quality flag indicating a very poor profile fit (as expected for a close visual binary), so we also derived masses using the same method but without using the 2MASS measurement (and only the \Gaia\ $G$-band magnitude), which yielded more conservative mass estimates of $0.346\pm0.027M_\odot$ and $0.331\pm0.024M_\odot$. We choose to adopt these more conservative estimates to avoid any possible systematics associated with the 2MASS data. 

We checked these results for consistency by fitting\cite{stassun2018} the SED of the two stars instead of empirical relations. Here, we fit only the resolved \Gaia\  $G$, $B_P$, and $R_P$ magnitudes.  We fixed the effective temperature of each M-dwarf to the values determined in the TICv8\cite{tic} ($T_{\rm{eff, A}} = 3521$ K  and $T_{\rm{eff, B}} = 3513$ K) since those were already based on the resolved Gaia $G_{BP} - G_{RP}$ colors, and determined the bolometric flux of the two stars using the \Gaia\ parallax. We determined the radii of the two stars to be $R_{\star,A} = 0.35 \pm 0.02 \rsun$, and $R_{\star,B} = 0.34 \pm 0.02 \rsun$. Converting from radii to masses using relations between the mass/radius of M-dwarfs and their absolute K-band magnitudes\cite{mann:2015, mann2019} yields $M_{\star,A} = 0.335 \pm 0.024 \msun$, and $M_{\star,B} = 0.322 \pm 0.023 \msun$. These results are in good agreement with our adopted masses.

\subsection{Triple System Orbit Analysis}
We investigated the orbits of the three stellar components in the WD 1856 system of WD 1856 and G 229-20 A/B about the system's center of mass. \Gaia\ DR2 measured highly precise positions and proper motions for the three stars, so we used the Linear Orbits for the Impatient (LOFTI\cite{lofticode}) algorithm\cite{pearce20} to derive orbital constraints from these observations. Given input proper motions, positions, radial velocities (if available), and masses of the stellar components, LOFTI uses rejection sampling\cite{blunt} to determine probability distributions for different orbital parameters. 

We ran LOFTI to determine parameters for the orbit of WD 1856 and G 229-20 A/B about the system's center of mass. For the latter, we approximated G 229-20 A/B as a point mass. We used the masses determined in our earlier analysis, and ran LOFTI until the rejection sampling algorithm had accepted 50,000 possible orbits. We found that the outer orbit is likely viewed close to face on (inclination $i = 22^{+11}_{-11}$ degrees) and may be modestly eccentric ($0.30\ensuremath{^{+0.19}_{-0.10}}$). The semimajor axis is $a=1500\ensuremath{^{+700}_{-240}}$ AU, and the separation between WD 1856 and the center of mass of G 229-20 at closest approach is $a\,(1-e)= 1030\ensuremath{^{+130}_{-55}}$ AU. 

We also ran LOFTI to determine parameters for the orbit of G 229-20 A and B about each other. Again, we ran the rejection sampler until we accumulated 50,000 samples in our posterior probability distribution.  G 229-20 A and B orbit with a semimajor axis $a = 58\ensuremath{^{+54}_{-16}}$ AU and have a separation of $a\,(1-e)= 39\ensuremath{^{+27}_{-20}}$ AU at their closest approach.  The eccentricity of the orbit is not well constrained, with $e<0.63$ (95\% confidence) and the posterior probability distribution for the inclination peaks near 50 degrees ($i = 51\ensuremath{^{+11}_{-17}}$ degrees).

\subsection{Transit Analysis}
\begin{figure}[htbp!]
\centering
\includegraphics[width=6.5in]{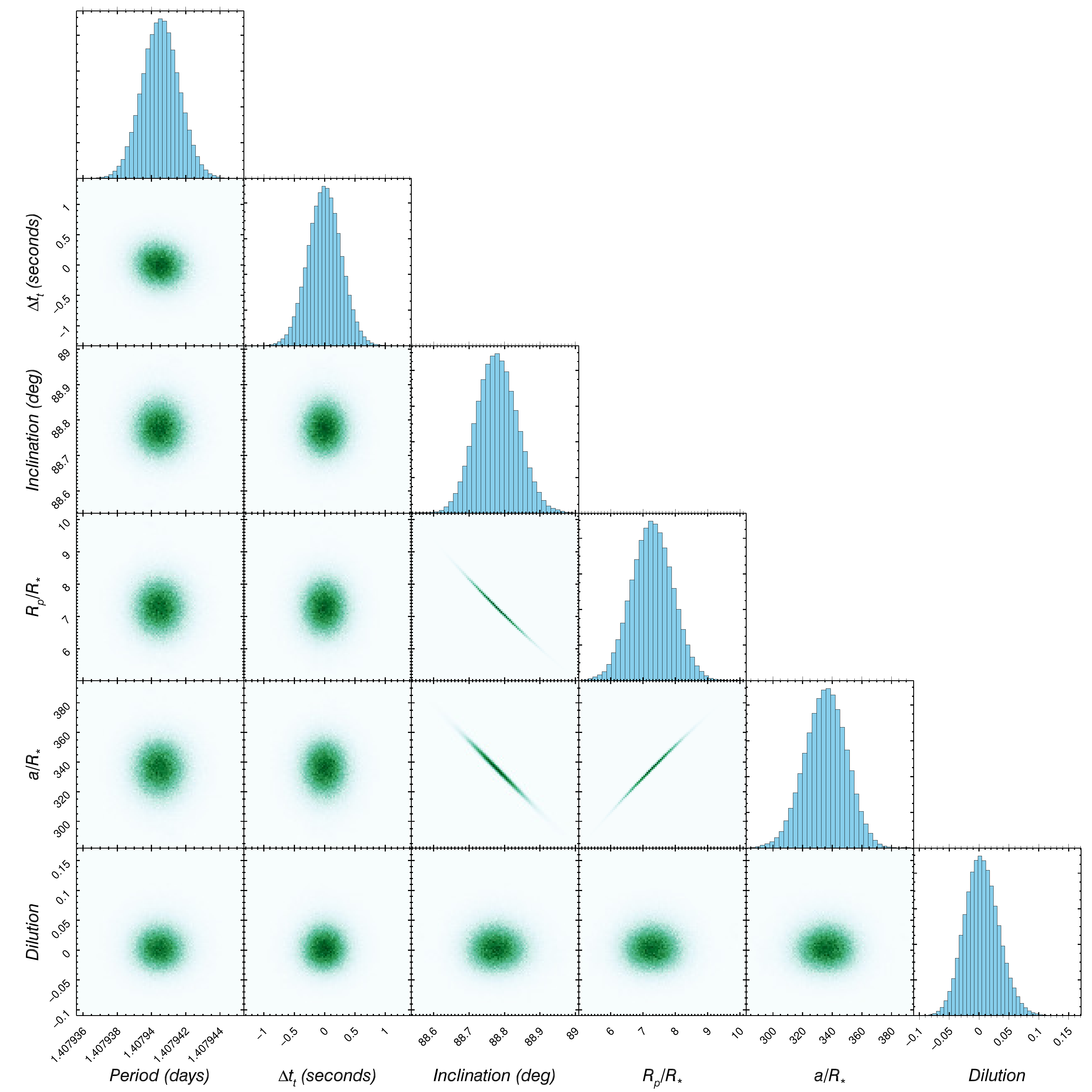}
\caption{\bedittwo{Posterior probability distributions of transit parameters.} This ``corner-plot'' shows correlations between pairs of parameters in our MCMC transit fit (with circular orbits enforced) and histograms of the marginalized posterior probability distributions for each parameter. For clarity, we have plotted correlations with the inclination angle $i$ instead of the fit parameter $\cos{i}$ and subtract the median time of transit ($t_t$). The orbital inclination $i$, scaled semimajor axis $a/R_\star$, and planet/star radius ratio $R_p/R_\star$ are strongly correlated due to the grazing transit geometry but constrained by the prior on stellar density. We do not include rows for the GTC and \Spitzer\ photometric jitter terms because these are nuisance parameters which showed no correlations with the other physical parameters.}
\label{corner1}
\end{figure} 

We determined the best-fit values and uncertainties on the transit parameters and the flux of WD 1856 b at 4.5 microns with a simultaneous MCMC analysis of the GTC and \Spitzer\ light curves.  We first selected a small portion of both the \Spitzer\ and GTC light curves near the observed transits; we used \Spitzer\ data collected at times $2458834.27\leq {\rm BJD}\leq 2458834.30$ and GTC data from $2458779.369\leq {\rm BJD}\leq 2458779.382$ (after converting the GTC timestamps to BJD\_TDB\cite{eastmantimes}). For convenience, we down-sampled the two-second-cadence \Spitzer\ light curve by a factor of 5 to match the 10-second cadence of the GTC light curve points. We divided the \Spitzer\ and GTC data  by the median out-of-transit flux measurement to set the out-of-transit flux level to 1. We estimated uncertainties on each point in the light curves by multiplying a value for the out-of-transit scatter (from the standard deviation of the normalized out-of-transit points) by the square root of each flux value. 

We fit the transits with \bedit{exact} analytic transit light curve models\cite{MandelAgol:2002} \bedit{for stars with quadratic limb darkening laws} coupled to a code for solving Kepler's equation\cite{Eastman:2013} (for fits with nonzero eccentricity). We oversampled the model light curves by a factor of 6 and integrated to account for the 10-second exposure time of both the GTC observations and our binned \Spitzer\ observations. We fixed the limb darkening parameters for the white dwarf to values calculated from model atmospheres. For our GTC $g^\prime$-band observation we used coefficients specifically calculated for white dwarfs \bedit{by Gianninas et al. (2013\cite{Gianninas13}). The Gianninas coefficients ($u_1 = 0.05$, $u_2 = 0.52$) closely match coefficients independently calculated by Claret et al. (2019\cite{Claret2019}, $u_1 = 0.07$, $u_2 = 0.46$)}. For our \Spitzer\ observation we used coefficients from models of main sequence stars with the same effective temperature\cite{claret2011} \bedit{($u_1 =0.0$, $u_2 = 0.15$)}.  We modeled WD 1856 b's flux contribution (if any) to the \Spitzer\ light curve by fitting for a dilution term $d \equiv F_{\rm WD\,1856\,b}/F_{\rm WD\,1856}$. We calculated and re-normalized the \Spitzer\ transit model $M_{S}(t)$ from the un-diluted transit model $M(t)$:

\begin{equation}
    M_{S}(t) = \frac{M(t) + d}{1 + d}
\end{equation}

At each MCMC link, we subtracted the transit models from the GTC and \Spitzer\ light curves, fit a quadratic polynomial to the residual light curves, and added this polynomial curve to the transit model. This step marginalizes over any possible trends and normalization errors in the two light curves. We fit for two additional photometric error terms (one for GTC and one for \Spitzer) added in quadrature to our calculated uncertainties and imposed a \bedit{Gaussian} prior on the density of WD 1856 centered at 324,000 \gcc\ with a width of 54,000 \gcc\ based on our stellar parameters. \bedit{Our knowledge of the stellar density lets us calculate WD 1856 b's average orbital speed via Kepler's third law (see Seager \& Mall\'en-Ornelas 2003\cite{SeagerMallenOrnelas2003}) and link the transit duration (a direct observable quantity) to the planet candidate's radius. This information, along with a constraint on the transit impact parameter from the maximum depth of the transit, helps the MCMC converge to a well-behaved solution. }

The transit of WD 1856 is grazing, so even when imposing a prior on the white dwarf's stellar density, the radius of the transiting object is almost completely degenerate with the object's orbital speed at the time of transit. We therefore performed one fit assuming a circular orbit and another fit allowing for orbital eccentricity. When we assumed circular orbits, we fit for 10 free parameters: orbital period, time of transit, cosine of the orbital inclination ($\cos{i}$), scaled semimajor axis ($a/R_\star$), planet/star radius ratio ($R_p/R_\star$), photometric jitter terms for both the \Spitzer\ and GTC light curves, and the \Spitzer\ dilution parameter $d$. Other than our prior on stellar density (which mostly affects $a/R_\star$), we used uniform priors with bounds $(-\infty, \infty)$ on all parameters except for the jitter terms, $a/R_\star$, $R_p/R_\star$, which we restricted to $[0,\infty)$, and $\cos{i}$, which we restricted to $[0,1]$. We did not impose a prior to force the dilution parameter to be positive to avoid a Lucy-Sweeney-like\cite{lucysweeney} bias. We explored parameter space with an affine invariant MCMC sampler\cite{goodmanweare} with 50 walkers evolved for 200,000 steps (discarding the first half for burn-in). 

For our fits allowing eccentric orbits, we changed our parameterization to speed the MCMC convergence. Instead of exploring parameter space in $\cos{i}$, we defined a new parameter $\delta \equiv R_p/R_\star - b$, where $b = a/R_\star \cos{i}$ is the transit impact parameter to avoid a strong correlation between $R_p/R_\star$ and $b$. We also fit for combinations of eccentricity $e$ and argument of periastron $\omega$ ($\sqrt{e}\sin{\omega}$ and $\sqrt{e}\cos{\omega}$) for a similar reason. We imposed a physical cutoff for high eccentricity orbits; at each link, we calculated WD 1856 b's instantaneous Roche lobe radius\cite{kopal1959close} at periastron $R_L$: 
\begin{equation}
    R_L \approx 0.46\,(1-e)\,a\left( \frac{M_p}{M_\star}\right)^{\sfrac{1}{3}}
\end{equation}
assuming a planet mass $M_p = 15$ \mjup\ (see below). We discarded any links where the planet's size exceeded this radius, \bedit{which prevented the fit from diverging towards high eccentricities and large companion radii}. Even with these modifications, the eccentric fit was much slower to converge; we evolved 50 walkers for \bedit{8,000,000 links, discarding the first 5,000,000 to remove the burn-in phase and save disk space}. \bedit{Correlations between selected parameters for both the circular and eccentric fits are shown in Extended Data Figures  \ref{corner1} and \ref{corner2}.}

Both fits showed that WD 1856 b is a roughly Jupiter-sized object. If its orbit is circular, WD 1856 b has a radius $R_p$ = \rplb \urplb \rearth; if eccentric orbits are allowed, the uncertainty on the radius is significantly larger: $R_p$ = \rplbecc \urplbecc \rearth. \bedittwo{Radii smaller than about 7 \rearth\ are strongly ruled out in both cases, so the companion cannot be another white dwarf.} Our fits also revealed that the transit depth at 4.5 micron wavelengths is nearly identical to the optical transit depth. We measure the \Spitzer\ dilution parameter $d = 0.004 \pm\ 0.029$. Evidently, the flux of WD 1856 b is only a small fraction of the white dwarf itself at 4.5 microns. This places strong constraints on the temperature (and therefore mass) of WD 1856 b, as described below. 

\bedit{In principle, using inaccurate limb darkening coefficients in our fits can adversely affect our measurement of the dilution coefficient and planet radius. We tested the robustness of our results to such errors by running additional MCMC fits where the limb darkening coefficients were free parameters constrained by basic physical priors\cite{kippingld}. We ran three separate fits: one where the \Spitzer\ limb darkening coefficients were restricted to likely values ($u_1<0.2$, $u_2<0.3$)\cite{claret2011} and the GTC coefficients were fixed to model values; one with the \Spitzer\ coefficients free and the GTC coefficients fixed to the model values; and one where both the GTC and \Spitzer\ limb darkening coefficients were free. Our results are insensitive to the limb darkening coefficients; our fit with the \Spitzer\ coefficients restricted to ($u_1<0.2$, $u_2<0.3$) and GTC coefficients fixed to model values gave statistically identical results to our baseline fit.  Even when both the \Spitzer\ and GTC coefficients were allowed to freely vary, the dilution parameter and $R_p/R_\star$ shifted by only 0.2$\sigma$ and 0.4$\sigma$, respectively.    }%values independently corroborated by two different model grids

\begin{figure}[ht!]
\centering
\includegraphics[width=5in]{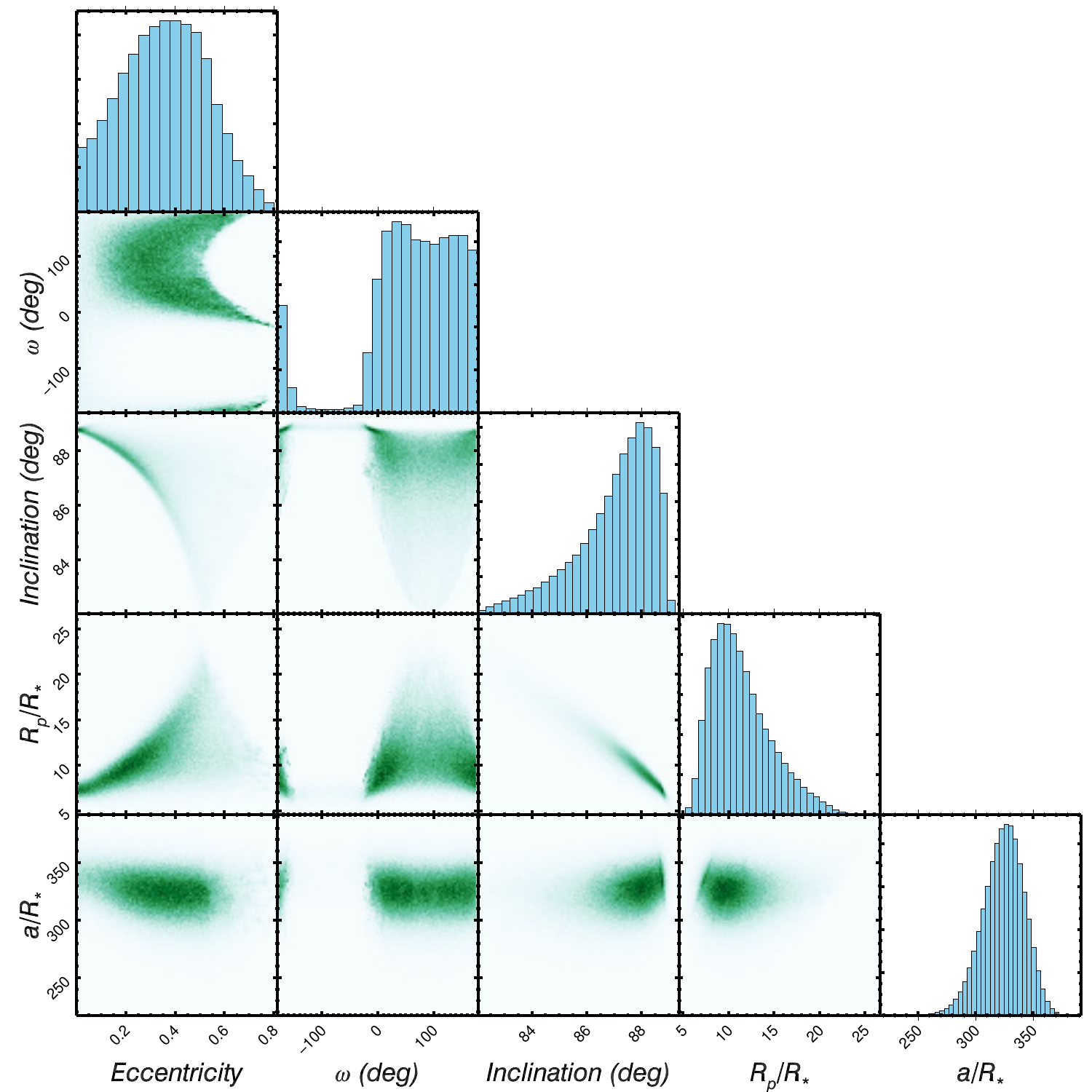}
\caption{\bedittwo{Posterior probability distributions of transit parameters when eccentric orbits are allowed.} This ``corner-plot'' shows correlations between pairs of parameters in our MCMC transit fit (allowing eccentric orbits) and histograms of the marginalized posterior probability distributions for each parameter. This plot shows a subset of the parameters that correlate with the orbital eccentricity. For clarity, we have plotted correlations with the eccentricity $e$, argument of periastron $w$ and orbital inclination $i$ instead of the fit parameters $\sqrt{e}\cos{\omega}$, $\sqrt{e}\sin{\omega}$, and $\delta$.}
\label{corner2}
\end{figure}

\subsection{Companion Mass Limit}
We quantified the constraints placed by our \Spitzer\ observations using brown dwarf/giant planet evolutionary and atmosphere models. From our measurement of $d = F_{\rm WD\,1856\,b}/F_{\rm WD\,1856}$ at 4.5 microns from our transit fits, and our measured total flux of WD 1856 and WD 1856 b at 4.5 microns (173 $\pm$ 10 \microjanskys), we calculate the flux of WD 1856 b at 4.5 microns:
\begin{equation}
    F_{\rm WD\,1856\,b} = d\,F_{\rm WD\,1856} = \frac{F_{\rm total}}{1+\sfrac{1}{d}} = 0.7 \pm 4.9 \microjanskys
\end{equation}
When we exclude all unphysical solutions where $d < 0$, we calculate 68\%, 95\% and 99.7\% upper limits on $F_{\rm WD\,1856\,b}$ at 4.5 microns that are 5.2, 10.2, and 15.5 \microjanskys, respectively. We emphasize that this limit on WD 1856 b's flux at 4.5 microns is \textit{model independent} and does not rely on our white dwarf stellar parameters or SED fit.

We used the Sonora grid\cite{sonora} of cloud-free solar metallicity brown dwarf/giant planet models to relate the thermal flux at 4.5 microns to atmospheric parameters like effective temperature and surface gravity. We interpolated the predicted thermal flux in IRAC Channel 2 from the Sonora atmosphere models onto two sets of evolutionary models: the underlying models used in the Sonora atmosphere calculations, and a more densely-sampled grid of models\cite{nelson2018} produced using the Modular Experiments in Stellar Evolution (MESA) code. We found that the two evolutionary grids gave nearly identical results, and adopted the MESA models given their denser sampling. 

The MESA brown dwarf models predict the properties of objects with masses from 2.1 \mjup\ to 104 \mjup\ over 20 Gyr of evolution and are sampled at a total of 329,732 points in the mass/age plane. We compared the predicted 4.5 micron flux for each of these model points to determine the allowed brown dwarf masses given our constraints. We assume that WD 1856 b must be at least as old as the white dwarf's cooling age (roughly 5.85 Gyr) and cannot be older than the age of the universe (13.8 Gyr), so we ignore any model points outside this age range. We found that for the oldest (13.8 Gyr) possible brown dwarfs/giant planets, we constrain the mass to be less than 11.1 \mjup\ at 68\% confidence (1$\sigma$), 13.8 \mjup\ at 95\% confidence (2$\sigma$), and 16.1 \mjup\ at 99.7\% confidence (3$\sigma$). The object's temperature must be below (250 K, 290 K, 320 K) at (1$\sigma$, 2$\sigma$, 3$\sigma$) confidence. 

The tail of WD 1856 b's allowed mass distribution straddles the 13 \mjup\ deuterium burning limit traditionally used to distinguish giant planets and brown dwarfs\cite{boss2007, saumonmarley08,1985Natur.316...42N}. However, using the deuterium burning limit to distinguish planets from brown dwarfs is imprecise. There is likely no specific mass above which deuterium burning takes place in brown dwarfs;\cite{deuterium} instead the limit likely spans a range from about 11-16 \mjup\ (depending on the object's composition and how one defines the onset of deuterium burning). It may also be more appropriate to divide planets and brown dwarfs by their formation histories\cite{chabrier, bowler20}. Given the lack of a clear division between planets and brown dwarfs, we refer to WD 1856 b as a planet candidate until future observations can place stronger constraints on its mass.

\bedit{These upper limits on WD 1856 b's mass are model dependent, so we tested how they change when we use different model grids and assumptions. We repeated our calculation using the new ATMO 2020 evolutionary and atmospheric models\cite{atmomodels}. Since these models were only calculated to an age of 10 Gyr, we compared the 1$\sigma$, 2$\sigma$, and 3$\sigma$ upper mass limits with those for for 10 Gyr objects with the Sonora/MESA models. We found good agreement in the mass upper limits between the two models (within about 2 \mjup, with ATMO 2020 models yielding a lower 1$\sigma$ mass limit and a higher 3$\sigma$ mass limit due to stronger dependence of 4.5 $\mu m$ flux on mass). We also tested the effects of non-equilibrium chemistry, which can be important for cold brown dwarfs\cite{miles20}, using the ATMO 2020 models.  Even strong disequilibrium chemistry ($\log{K_{zz}}\sim 6.5$) had a minimal effect on our mass limits.} 

\bedit{The effect of clouds on our mass limits is more difficult to quantify. In general, the presence of clouds slows the cooling of brown dwarfs and giant planets, so objects with clouds should generally remain hotter and more luminous throughout their evolution\cite{saumonmarley08}. However, when clouds are present, they can significantly change the object's spectrum and tend to decrease the flux in the 4.5 $\mu m$ band\cite{morley2018}. Water clouds are expected to form in giant planets and brown dwarfs cooler than about 375 K\cite{morley2014}, so in the case of WD 1856 b, these two effects will likely compete. Future modeling should more fully reveal which effect dominates.}

\subsection{Age of the WD 1856 system}
Because giant planets and brown dwarfs cool as they age, our mass limits are stronger for younger systems. We therefore attempted to place additional constraints on the total system age in addition to the white dwarf cooling age (age $\gtrsim 5.85$ Gyr) and the age of the universe (age $< 13.8$ Gyr) . One possible way to measure the age of a white dwarf is to add the white dwarf's cooling age to the estimated main sequence lifetime of its progenitor star using a white dwarf initial/final mass relation. Unfortunately, two factors make it difficult to estimate the progenitor's age. First, the white dwarf initial/final mass relations assume the star evolved as an isolated single star and did not undergo mass/transfer or a common envelope phase. As we show below, though it is difficult, it is perhaps not impossible that WD 1856 b reached its current orbit by this mechanism. Second, a white dwarf progenitor's lifetime is a sensitive function of the white dwarf's final mass; a 50\% increase in a white dwarf's mass from 0.5 \msun\ to 0.75\msun\ corresponds to a 275\% increase in the progenitor's mass from 0.8 \msun\ to 3 \msun\ and a corresponding factor of $\sim$20 decrease in the star's main sequence lifetime (from $\sim$ 10 Gyr to $\sim 500$ Myr). With a mass of 0.52 \msun, the white dwarf initial/final mass relation favors a long-lived progenitor with a mass less than that of the Sun and a total system age at least 15 Gyr, older than the age of the universe. Since our white dwarf model spectra struggle to describe our observations (see above), we suspect that systematic errors in our estimate of WD 1856's mass likely explain the system's apparently unphysical age. If the true mass were closer to 0.6 \msun (only $\approx1.5\sigma$ away given our conservative uncertainties), this tension would disappear. We conclude that given these uncertainties, estimating WD 1856's progenitor's lifetime cannot give a reliable system age. 

\begin{figure}[h]
\centering
\includegraphics[width=5.5in]{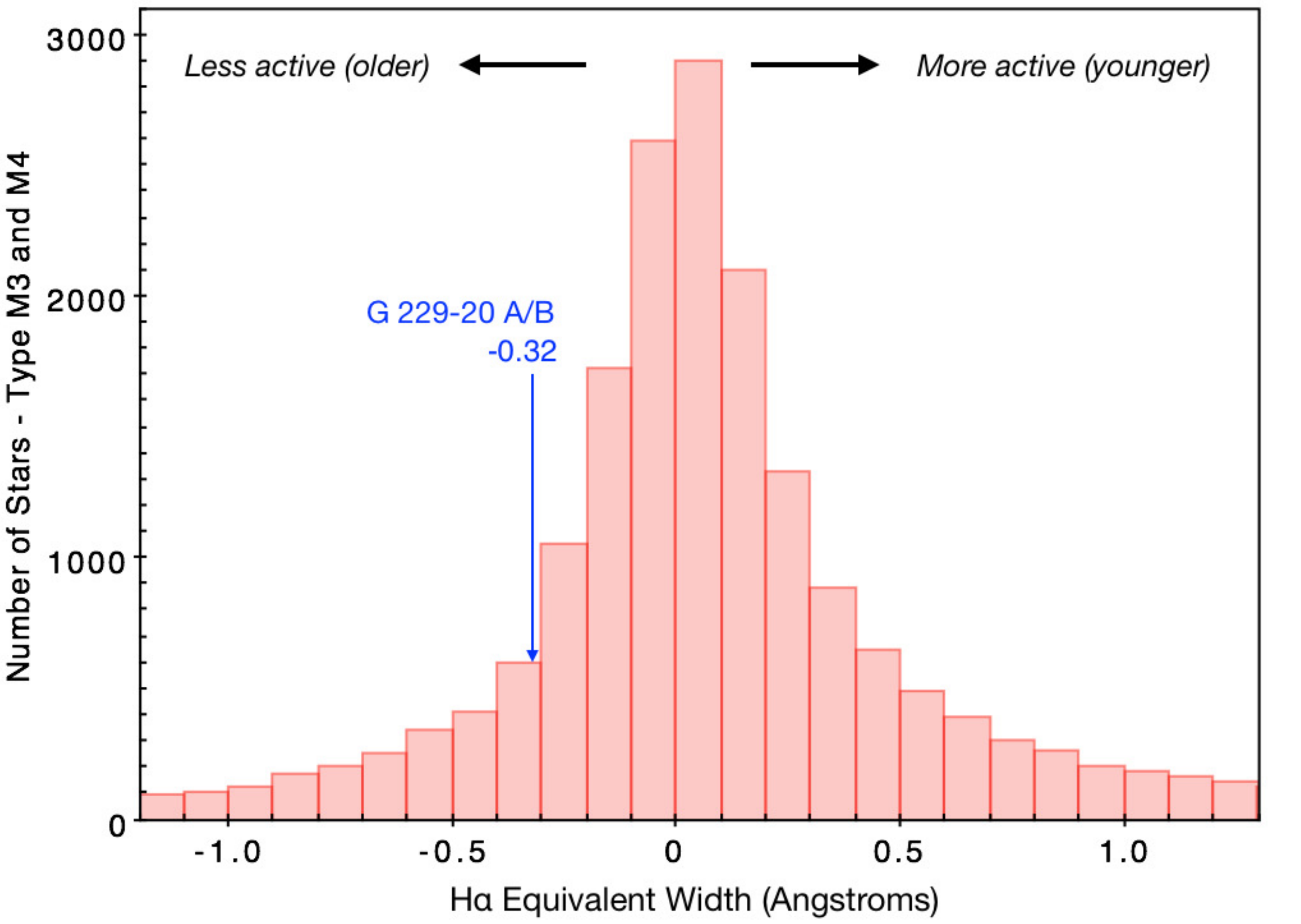}
\caption{\bedittwo{\halpha\ equivalent width for G 229-20 A/B compared to other nearby M-dwarfs.} The histogram shows the \halpha\ equivalent widths for large sample of M-dwarfs with similar spectral types from the Sloan Digital Sky Survey\cite{west11}. G 229-20 A/B (shown as a blue arrow) have lower than average \halpha\ equivalent width, but fall well within the distribution of field M-dwarfs.}
\label{halphaew}
\end{figure} 

We then shifted our attention to the binary M-dwarf pair G 229-20 A/B. Presumably these stars formed together with WD 1856's progenitor, and therefore should be the same age as WD 1856's planet candidate. It is notoriously difficult to determine the age of old ($\gtrsim$ 1 Gyr) field stars, and especially difficult for M-dwarfs, but there are some indicators which can broadly suggest an age for the system. We saw no evidence that the M-dwarfs are particularly young; the two stars do not have \halpha\ in emission, and light curves of the two stars from \TESS, the ASAS-SN survey, and the SuperWASP survey show no evidence for a rotational variability. This is unsurprising since we assume G 229-20 A and B must have formed before WD 1856 became a white dwarf about 5.85 Gyr ago. However, we also saw no evidence that G 229-20 A/B are particularly old. Like most typical field age M-dwarfs, the spectra G 229-20 A/B show a band of prominent Calcium Hydride (CaH) and Titanium Oxide (TiO) absorption features\cite{gizis} often characterized using the $\zeta_{TiO/CaH}$ parameter\cite{lepine2006, mann2013}; if G 229-20 A/B were old sub-dwarfs, we would expect $\zeta_{TiO/CaH}<0.8$, but the value is 0.93, consistent with most Solar-metallicity M dwarfs. G 229-20 A/B's \halpha\ equivalent width (a proxy for magnetic activity and therefore age\cite{newton2017}) is lower than average, but still well within typical ranges for field M-dwarfs\cite{west11} (see Extended Data Figure \ref{halphaew}).  We also investigated the system's galactic kinematics. Using the system's position, proper motion, and parallax from \Gaia\ DR2 along with our measured radial velocity (with an inflated uncertainty to account for the M-dwarfs' motion about the system barycenter), we calculated the system's 3D space motion to be (U,V,W) = (8.65$\pm$0.21, 40.4$\pm$1.8, -15.13$\pm$0.70) \kms\ with respect to the Local Standard of Rest (LSR\cite{lsr}).  We calculated the relative probabilities\cite{bensby14, carrillo2020} that the WD 1856 system is a member of the galactic thin disk, thick disk, or halo, and found that WD 1856 is most likely (93\%) a member of the thin disk, with only about a 7\% chance that it is part of the thick disk. Halo membership is strongly disfavored (4000:1 odds against). The mean age for stars in the thin disk is about 7-8 Gyr\cite{kilic2017} (with large spread), and the oldest stars in the thin disk are probably around 8-10 Gyr in age\cite{haywood2013, xiang}. Thick disk stars are about 1.5-2 Gyr older on average than thin disk stars, with a mean age of $\approx$9-10 Gyr\cite{kilic2017, sharma}. 

All in all, these lines of evidence point to a system that is fairly old, but not likely to be much older than about 10 Gyr. If we assume the system is no older than 10 Gyr, WD 1856 b's mass must be less than (9.4 \mjup, 11.9 \mjup, 13.6 \mjup) at (1$\sigma$, 2$\sigma$, 3$\sigma$) confidence.

\subsection{Common Envelope Evolution}
When WD 1856's progenitor star was on the main sequence, the companion WD 1856 b must have orbited farther from the progenitor than it does today, or it could not have survived the progenitor's red giant evolutionary phase. Here, we consider how WD 1856 b might have reached its current orbit close to WD 1856. One obvious possibility for placing a massive planetary object in a relatively close orbit with a white dwarf is common envelope \bedit{evolution}\cite{Paczynski1976, Webbink84, Pfahl03}. Nelson et al. (2018)\cite{nelson2018} investigated the likelihood that short-period, detached binaries containing a brown dwarf (or low-mass M-dwarf) companion in orbit with a white dwarf (or hot subdwarf) could have been formed via a common envelope (`CE') phase of evolution. They compiled a table of 25 binaries with orbital periods between 68 min and 4 hours and showed that the measured masses of the companions, which typically fall in the range of 50–100 \mjup, are not inconsistent with the predictions of CE evolution. There are some detached systems having orbital periods longer than 4 hours with companion masses in this range, but none that we are aware of with periods as long as that of WD 1856 (1.4 days). Nonetheless, we will now examine whether it is possible for a 15 \mjup~object (at the upper end of our allowed mass distribution) to eject the envelope of a low-mass giant and end up in an orbit as long as 1.4 days.

There are a number of different ways to formulate the initial-final orbital separation ($a_f-a_i$) during a CE phase based on conservation of energy.  The basic idea is to determine the final binary orbital separation once the low-mass companion has ejected the CE of the progenitor, in terms of the initial orbital separation of the primordial binary and its constituent masses.  More recent treatments of the energy formulation take into account the fraction of the internal energy used to eject the envelope, for example the recombination energy\cite{Zorotovic10,xu2010, 2011MNRAS.411.2277D, camacho}. 
Conservation of energy relates $a_f$ to $a_i$ as follows: 
\begin{equation}
\frac{GM_p M_e}{\lambda r_L a_i}=\alpha \left[\frac{GM_c M_s}{2 a_f}-\frac{G M_p M_s}{2a_i}\right],
\label{eqn:CE1}
\end{equation}
where $M_p$ and $M_s$ are the masses of the primordial primary (the WD progenitor) and the primordial secondary star (in this case the massive planet), respectively, and $M_c$ and $M_e$ are the masses of the core and envelope of the primary star\cite{Taam78, Webbink84,Taam92}. The parameter $\lambda^{-1}$ is a measure of the total gravitational binding energy of the envelope to itself and to the core of the primary star in units of $-GM_p M_e/R_p$, while $\alpha$ is an energy efficiency parameter for ejecting the common envelope.  The factor $r_L \equiv R_L/a_i$ is the dimensionless radius of the Roche lobe of the primary star when mass transfer commences.  If the internal energy (e.g., electron recombination) is taken into account, then either $\alpha$ or $\lambda$ may be considered to be larger than unity\cite{TaurisDewi,xu2010, camacho}.  

\begin{figure}[h]
\centering
\includegraphics[width=4.in]{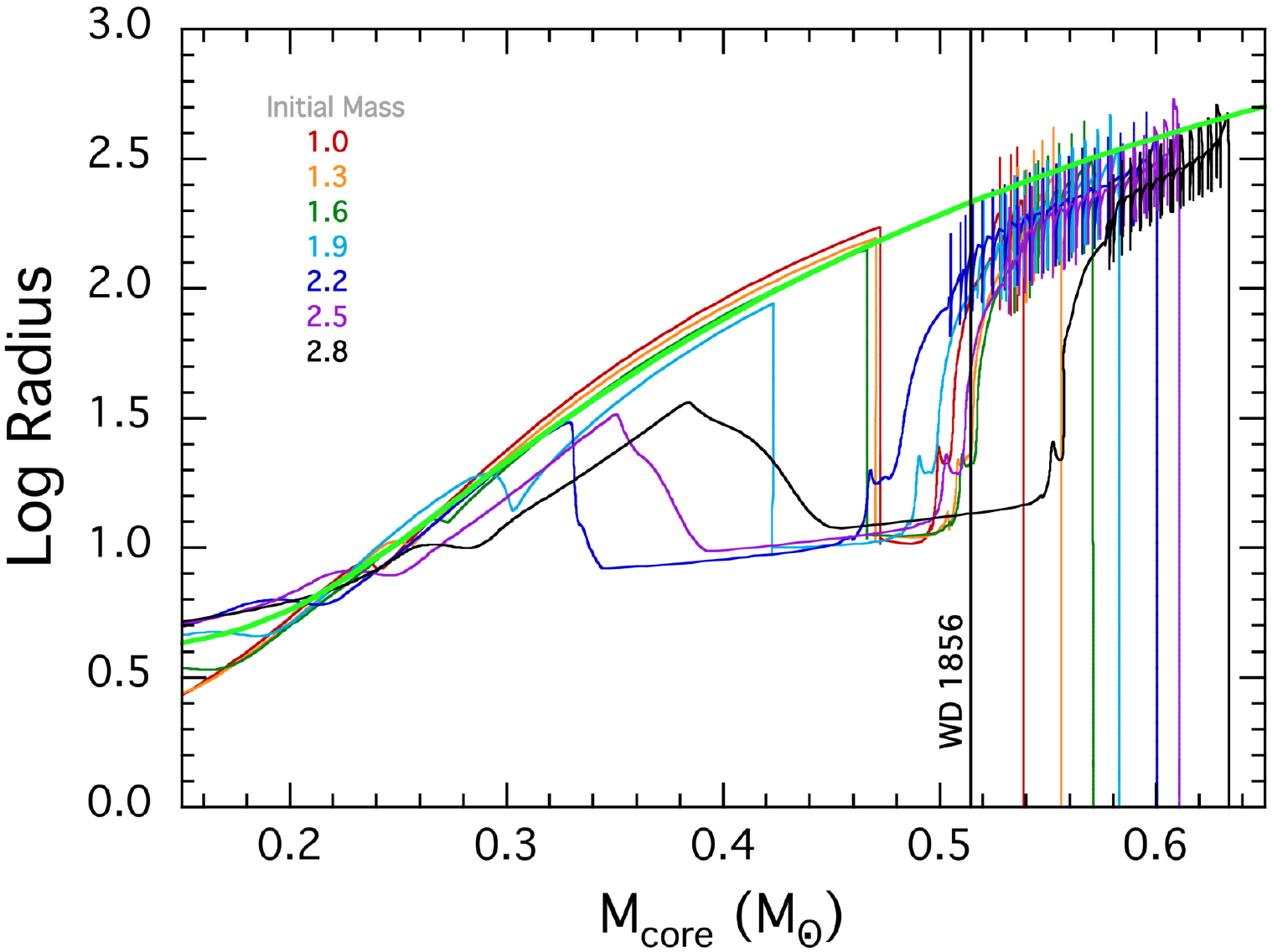}
\caption{\bedittwo{Theoretical relationships between the a star's radius and the mass of its core.} We show MIST\cite{mist} evolution tracks in the radius--core-mass plane for solar composition models with masses ranging from 1-2.8 $M_\odot$. The RGB phase is clearly identifiable for core masses between 0.2 and 0.47 $M_\odot$, while the thermal pulses on the AGB are readily recognized at higher core masses of $\gtrsim 0.5 \, M_\odot$.  The lime green curve is the analytic expression given by Eqn.~(\ref{eqn:r(mc)}). The vertical lines for each star mark the point where the envelope has been exhausted by the AGB wind. }
\label{fig:mist}
\end{figure} 

For the masses and separations relevant to the formation of the WD 1856 system, the second term in square brackets in Eqn.~(\ref{eqn:CE1}) is negligible compared to the first term (see Rappaport et al., 2015\cite{rappaport15} for a more detailed analysis).  Upon dropping that term, we find:
\begin{equation}
\frac{a_f}{a_i} \simeq \frac{\lambda \alpha r_L}{2} \left(\frac{m_c m_s}{m_e m_p}\right).
\label{eqn:CE2}
\end{equation}
where lower-case masses are implicitly expressed in solar units. In turn, this can be expressed as the ratio of final to initial orbital periods:
\begin{equation}
\frac{P_f}{P_i} \simeq \left(\frac{\lambda \alpha r_L}{2}\right)^{3/2} \left(\frac{m_c m_s}{m_e m_p}\right)^{3/2} \left(\frac{m_p+m_s}{m_c+m_s}\right)^{1/2}.
\label{eqn:CE3}
\end{equation}
Since the mass of the degenerate core of low-mass stars is closely related to the radius of the giant, it also follows that there is a relation between the orbital period and giant's core mass when mass transfer commences.  

We illustrate the $R(M_c)$ relation in Extended Data Figure~\ref{fig:mist}.  Here we show {MIST}\cite{mist} evolution tracks for solar metallicity stars in the radius--core mass plane. These are for 7 different initial stellar masses covering the range of 1.0 to 2.8 $M_\odot$. On the first red giant branch there is a common locus of upper limits to the radius, while on the Asymptotic Giant Branch (AGB) the same is true, with the main difference being the thermal pulses during which the radius varies substantially.  The lime green curve superposed on the plot is an analytic expression that represents fairly well the locus of upper limits -- which is where mass transfer to a companion star would first occur.  The expression 
\begin{equation}
R(m_c) \simeq 5.56 \times 10^4 \frac{m_c^{19/3}}{1+20 m_c^3+10  m_c^6} + 4 ~~~R_\odot
\label{eqn:r(mc)}
\end{equation}
(for 0.7 $\gtrsim \, M_c \gtrsim 0.15 \, M_\odot$) is modeled after Eqn.~(5) in Rappaport et al. (1995\cite{rappaport95}) and inferred from Eqn.~(12) in Kalomeni et al. (2016\cite{kalomeni16}), with some minor modifications.

The orbital period that corresponds to a primary with core mass $m_c$ and which is just filling its Roche-lobe with the secondary star is:

\begin{eqnarray}
P_i  & \simeq & 1.53 \times 10^6 f(m_{\rm c})^{3/2}  \frac{1}{r_L^{3/2}}\frac{1}{\sqrt{m_p+m_s}} ~~{\rm days} \label{eqn:PM} \\
{\rm with:} ~~~~f(m_{\rm c}) & \equiv & \frac{m_{\rm c}^{19/3}}{(1+20m_{\rm c}^3 +10 m_{\rm c}^6)}+f_0  \nonumber
\label{eqn:pinit}
\end{eqnarray}
where $f_0 = 7.2\times 10^{-5}$.  Here $r_L$ has the same meaning as in Eqns.~(\ref{eqn:CE2}) and (\ref{eqn:CE3}).

We now combine Eqns.~(\ref{eqn:CE3}) and (\ref{eqn:PM}) into a single equation for the post-common envelope period, $P_{\rm pce}$, and associate the system masses in Eqn.~(\ref{eqn:CE3}) with those we observe in WD 1856: $m_c \equiv m_{\rm wd}$, $m_s \equiv m_{\rm com}$, and $m_e \equiv m_p - m_c$, where the subscript ``com'' stands for the current companion to the WD which we believe is a gas-giant planet: 
\begin{equation}
P_{\rm pce}  \simeq  5.4 \times 10^5 (\lambda \alpha)^{3/2}  \frac{f^{3/2}(m_{\rm wd})}{\left(m_{\rm wd}+m_{\rm com}\right)^{1/2}}
 \left[\frac{m_{\rm wd}\,m_{\rm com}}{(m_p-m_{\rm wd}) m_p}\right]^{3/2}    ~~{\rm days}.  
\label{eqn:ppce}
\end{equation}
Note that the period of the post-CE system is a function only of the masses of the companion, the white dwarf, and its progenitor.

For the WD 1856 system we know $P_f = 1.4$ d, $M_{\rm wd} = 0.52 \, M_\odot$, and we will take $M_{\rm com} \simeq 0.015 \,M_\odot$ as an upper limit on the mass of the current companion object. Thus, we can use Eqn.~(\ref{eqn:ppce}) to find the required value of $\alpha \lambda$ as a function of the primary mass (progenitor of the WD):

\begin{equation}
\alpha \lambda  \simeq  1.5 \times 10^{-4} \, P_{\rm pce}^{2/3} \,  f(m_{\rm wd})^{-1} \frac{\left(m_{\rm wd}+m_{\rm com}\right)^{1/3}  (m_p-m_{\rm wd}) m_p } { m_{\rm wd}\,m_{\rm com} }  .
\label{eqn:alpha_lambda}
\end{equation}

Finally, in Extended Data Figure~\ref{fig:alpha_lambda_combined} we plot Eqn.~(\ref{eqn:alpha_lambda}) as a function of the mass of the primary progenitor star of the current WD.  From this figure we can see that for progenitor masses of 1, 2, and 3 $M_\odot$, values of the parameter $\alpha \lambda = 2.4, 15$, and 38 would be required to unbind the envelopes.  According to Xu et al. (2010\cite{xu2010}), the calculated values of $\alpha \lambda$, including internal energies are $\lesssim 0.4$, $\lesssim 2$, and $\lesssim 5$, respectively (when the stellar radii are in the relevant range of 100--250 $R_\odot$), considerably less than the values required for WD 1856 b to eject the primary star's envelope. Without invoking internal energy, it appears even more improbable that a 15 $M_J$ object could unbind the common envelope of the white dwarf's progenitor.

We explored whether WD 1856 b could have plausibly ejected a common envelope at any point in its progenitor's evolution by calculating the required $\alpha \lambda$ value from the MIST tracks directly. At each point in the MIST tracks where the primary star was expanding to engulf new regions of its solar system, we calculated the required $\alpha \lambda$ assuming an orbit for WD 1856 b such that the primary star was just filling its Roche lobe. We calculated the minimum $\alpha \lambda$ during three different intervals in the progenitor star's evolution: before the star reached the thermally pulsating AGB phase and began rapidly losing mass, before 30\% of the progenitor's envelope mass had been lost, and at any point in the star's evolution. Our values for $\alpha \lambda$ as a function of stellar mass and at different points in the progenitor's evolution are also shown in Extended Data Figure \ref{fig:alpha_lambda_combined}. Our curve of the minimum $\alpha \lambda$ prior to the AGB confirms the results from our analytic study: it is energetically difficult for WD 1856 b to eject the envelope while most of its mass is still in place. Even once 30\% of the envelope's mass is lost, it is still difficult to eject the envelope; typical $\alpha \lambda$ values of 1-10 indicate that WD 1856 b's gravitational potential energy is insufficient, but the envelope perhaps could be ejected if a large fraction of the envelope's internal energy contributed to its ejection. \bedit{By the} very end of the AGB phase, once about 50\%-60\% of the envelope's mass has been lost, the minimum $\alpha \lambda$ values become less than unity. \bedit{The observed population of post common envelope binaries suggests\cite{camacho} that towards the end of the AGB phase, $\lambda$ could be as high as 10, so it is possible that WD 1856 b could eject its progenitor's envelope (though the population also favors values of $\alpha \lesssim 0.3$). However, given the relatively small region of parameter space in which this mechanism could produce WD 1856 b's current orbit, we consider common envelope evolution less likely than the dynamical explanation outlined below. } 

\begin{figure}[h!] 
\centering
\includegraphics[width=6.0in]{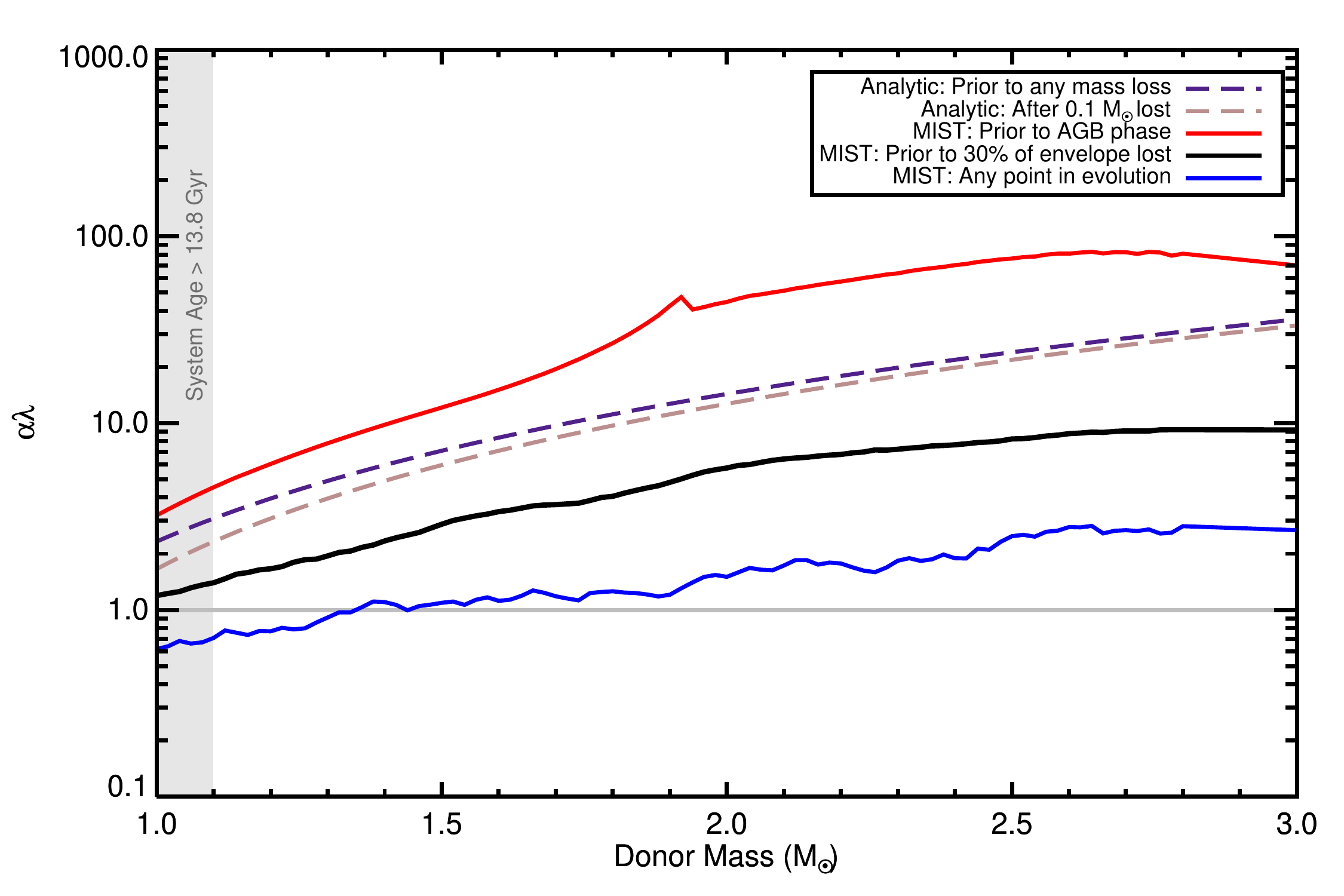}
\caption{\bedittwo{The minimum value of the efficiency parameter $\alpha \lambda$ required for WD 1856 b to form via common envelope as a function of the progenitor stellar mass. The two dashed curves show the minimum $\alpha\lambda$ values from our analytic calculation (Equation \ref{eqn:alpha_lambda}) required for a 15 \mjup\ object to eject the primary star's envelope. The purple dashed curve is taken directly from Equation \ref{eqn:alpha_lambda}, while the brown dashed curve results if the progenitor star has lost 0.1 $M_\odot$ in a stellar wind by the time of the common envelope.  The three solid curves curves show the minimum $\alpha \lambda$ computed directly from MIST tracks in three different situations: before the star reaches the AGB (red), before more than 30\% of the star's envelope mass has been lost (black), and at any point in the star's evolution, regardless of the mass lost (blue). Stars in the grey region at low masses evolve too slowly for the system to have left the main sequence more than 5.85 Gyr ago and are not viable solutions. For values of $\alpha \lambda > 1$ (horizontal grey line), one must invoke the internal energy of the star to help unbind the envelope during the common envelope phase.  Before mass is lost during the AGB phase, it is difficult for WD 1856 b to eject the common envelope, but it is possible WD 1856 b could have ejected its progenitor's envelope if the common envelope phase began after the progenitor reached the AGB. We have smoothed the lower two curves to remove some unphysical scatter likely due to numerical artefacts in the model grids.}}
\label{fig:alpha_lambda_combined}
\end{figure}

For planets that might manage to eject the envelope of the WD progenitor, at least in principle, there are some other perils that may await it. Passy et al. (2012\cite{2012ApJ...759L..30P}) examined whether planets and brown dwarfs would be disrupted by ram pressure during their passage through the dense inner envelopes of the giant during the common envelope phase.  They conclude that brown dwarfs and Jovian-mass objects (including a $10 \, M_J$ planet) are not likely to lose significant mass during their passage, whereas lower-mass planets could well be destroyed.  Bear \& Soker (2011\cite{bearsoker}) studied the mass loss of planets that might survive the common envelope, only to find themselves in the intense radiation of the nascent white dwarf \bedit{(see also Schreiber et al. 2019\cite{Schreiber2019})}. Bear \& Soker (2011\cite{bearsoker}) concluded that, while lower mass planets might be obliterated by evaporation, Jovian planets and those of higher mass might well survive to the point where the WD has cooled sufficiently for planetary evaporative losses to become insignificant.  Thus, if WD 1856 b had somehow been able to successfully eject the envelope of its progenitor, it might then survive the subsequent heating by the very hot white dwarf. However, we caution that these conclusions are very dependent on the assumed input physics of the models.

\subsection{Dynamical Formation}
Given the difficulty explaining WD 1856 b's current orbit with common envelope evolution we investigated other ways to form the system. Here, we consider whether WD 1856 b could have reached its current orbit as a result of dynamical scattering after WD 1856's progenitor evolved into a white dwarf. This \bedit{framework} has two main components: (i) perturbing WD 1856 b into a high-eccentricity orbit with a close periastron passage and (ii) dissipating the orbital energy to shrink the planet's semimajor axis and shorten the orbital period to 1.4 days. We consider these two processes separately. 

\textit{Generating a short periastron distance for WD 1856 b:} Since WD 1856 b must have formed and evolved far away ($\gtrsim$ 1 AU) from WD 1856's progenitor star, we explored whether dynamical processes can perturb a planet with a semimajor axis of roughly 1-2 AU into a highly eccentric orbit with a periastron distance of only a few solar radii. First, we considered whether the gravitational influence of WD 1856 b's M-dwarf companions (G 229-20 A/B) could excite a high eccentricity in WD 1856 b's orbit via the Kozai\cite{kozai}-Lidov\cite{lidov} effect\cite{stephan17}. We ran a small set of N-body simulations using \texttt{Mercury6}\cite{mercury6} with the four known bodies in the WD 1856 system, initialized with WD 1856 b in a circular orbit with a distance of 1-2 AU about WD 1856, and with G 229-20 A/B orbiting at a distance of about 1000 AU, consistent with the result of our LOFTI orbit fits (described above).  Under these conditions (and when the mutual inclination between the orbits of WD 1856 b and G 229-20 A/B is large enough), G 229-20 A/B do induce Kozai-Lidov cycles in WD 1856 b's orbit, but the timescales are slow ($\gtrsim$ 100 million years) and the amplitudes of the eccentricity oscillation are low (e $\sim$ 0.1). Although different values of initial conditions (including the eccentricities of both orbits) and mutual inclinations may alter the specific amplitudes and timescales of Kozai-Lidov oscillations, we conclude that it is difficult for G 229-20 A/B to excite WD 1856 b's orbit to $e\gtrsim 0.99$ eccentricity and close periastron passages.

Even if G 229-20 A/B could not have decreased WD 1856 b's periastron distance by exciting its eccentricity, it is possible that additional (undiscovered) bodies in the system could have.  Previous work \cite{debes2002, verasunpacking} has shown that systems of multiple planets residing exterior to the red giant expansion radius (but in a relatively well-packed configuration) can remain dynamically stable until after the WD has formed and begun cooling, then experience potentially violent instabilities.  Veras \& Gansicke (2015\cite{verasunpacking}) found that increasing the number of planets in their simulations resulted in more extreme dynamical evolution, including periastron passages as close as that of WD 1856 b. We ran an additional set of N-body simulations to confirm that the pattern seen by  Veras \& Gansicke (2015) holds true for systems with giant planets like WD 1856. Again, we used \texttt{Mercury6} to calculate the evolution of multi-planet systems. We initialized our simulations with up to four planets in closely packed orbits, with equal masses to WD 1856 b. Though our simulations are not an exhaustive exploration of parameter space, they do confirm that in multi-planet systems, violent dynamical instabilities can lead to planets being ejected from the system, sent onto a collision course with the white dwarf, or into orbits with small periastron distances.

\textit{Dissipating orbital energy and shrinking the semimajor axis:} If WD 1856 b had been perturbed into a highly eccentric orbit with a close periastron passage, it must have dissipated much of its orbital energy to end up with a 1.4 day period like we see today. We investigated whether tidal effects could dissipate WD 1856's orbital energy quickly enough to nearly circularize the planet's orbit in the roughly 5.85 Gyr cooling age of the white dwarf. Because WD 1856 is very small and dense, any tides raised on the white dwarf by the planet will be small and have negligible dissipative effects. Instead, any tidal dissipation in WD 1856 b's orbit must be due to tides raised on the planet by its star. 

The problem of tidally dissipating orbital energy for planets in highly eccentric orbis around white dwarfs has previously been studied by Veras and Fuller (2019a\cite{Veras2019a} and 2019b\cite{Veras2019b}). They calculated the total time needed to circularize a highly eccentric orbit as the sum of two different tidal regimes: a chaotic tidal regime at high eccentricities (e $\gtrsim 0.95$), where dissipation is dominated by the exchange in energy between the orbit and internal modes, and a classic tidal regime, at $e \lesssim 0.95$, where dissipation is dominated by equilibrium tides. Veras and Fuller calculate timescales for the completion of the chaotic tidal regime for gas giant planets and find typical values between 1 and 100 million years -- we conservatively choose a timescale at the high-end of their estimates for the WD 1856 system. We then estimated the time needed for the system to circularize from $e\approx0.95$  via equilibrium tides with:  
\begin{equation}
    t_{circ} = \frac{6 a^{5} Q_{p} m_{p}}{63  n_{p} k_{p} m_{*} R_{p}^{5}},
\end{equation}
where  $a$ is the planetary semimajor axis, $Q_{p}$ is the planetary tidal quality factor, $m_{p}$ the planetary mass, $n_{p}$ the planetary mean motion (related to the orbital period $P$ by $n=2\pi/P$), $k_{p}$ the planetary Love number, $m_{*}$ the stellar mass, and $R_{p}$ the planetary radius\cite{Goldreich1966}. Plugging in parameters for the WD 1856 system, and assuming WD 1856 b has Jupiter's mass, radius, and $Q/k_p$ (estimated\cite{Lainey2009} to be $Q_J/k_{p,J}\approx 10^5$), we estimate a tidal circularization timescale of about 2 Myr. Larger planet masses (5-10 \mjup) and more conservative estimatates of $Q/k_p$ up to $10^7$ should still circularize within the white dwarf's cooling age. All together, the timescale for tidal circularization of WD 1856 b's orbit is comfortably less than the system's age. 

We note that these processes could just as easily be applied to smaller planets than WD 1856 b. Packed systems of Earth-mass planets should exhibit the same dynamical instabilities that can drive close periastron distances for giant planets\cite{verasunpacking}, and tidal circularization should be even more efficient for rocky Earth-sized planets than gas giants like WD 1856. We estimate that tides raised on an Earth-sized planet should dissipate its orbital eccentricity within about 500,000 years. This formation pathway could potentially lead to the production of habitable-zone rocky planets\cite{agol2011}. Old white dwarfs cool slowly and could provide a relatively stable radiation environment for billions of years\cite{kozakis18}; we estimate that WD 1856 b's current orbital location was in the circumstellar habitable zone for almost 3 Gyr. WD 1856 b may demonstrate a mechanism that can lead to a second generation of habitability in a planetary system.

\subsection{Other theories}
We also explored other mechanisms that might be able to lead to WD 1856 b's current orbital configuration. We consider these other mechanisms less likely since they require either finely-tuned or \textit{a priori} unlikely initial conditions to succeed, but mention them for completeness. 

\textit{Close Stellar Encounters: } WD 1856 may have been perturbed from its initial, long-period orbit by a close flyby with another star. We estimated the most likely distance of closest approach $D_{\rm closest}$ between WD 1856 and another star during its 5.85 Gyr cooling age: 

\begin{equation}
    D_{\rm closest} \sim (\pi\,v\,t_{\rm cool}\,n)^{-\sfrac{1}{2}}
\end{equation}

\noindent where $v$ is the typical stellar velocity in WD 1856's vicinity ($\approx$ 60 \kms), $t_{\rm cool}$ is the cooling age (5.85 Gyr), and $n$ is the number density of stars in WD 1856's vicinity. We estimated $n$ using the fact that there are about 6000 stars within 25 parsecs of the sun from \Gaia\ DR2, giving a density of about 0.1 star per cubic parsec. We find $D_{\rm closest}\sim 600$ AU, so likely within its cooling lifetime, another star has passed by within the orbit of G 229-20 A/B. However, a much closer approach than 600 AU would be required to perturb WD 1856 b from a $\sim$1-10 AU orbit to a close periastron passage, and the probability $p$ of such a close approach decreases as $p\propto D_{\rm closest}^{-2}$ . 

\textit{Dynamical Instabilities from Galactic Tides:} \bedit{Bonsor \& Veras (2015\cite{bonsorveras}) suggested that galactic tides could perturb the orbit of a wide white dwarf binary and lead to a close approach billions of years after the system's formation. This mechanism could provide a trigger for dynamical instabilities in old white dwarf systems. In principle, such a mechanism could be important to the formation of WD 1856 b given the old system age and the presence of wide visual companions. Bonsor \& Veras (2015\cite{bonsorveras}) calculate that for galactic tides to be important on timescales of a few Gyr, the semimajor axis must be greater than about a few thousand AU and the wide binary orbit must be highly inclined with respect to the galactic plane (that is, the the pole of the orbit must be near the plane). Our fit to the WD 1856/G 229-20  orbit with LOFTI gives a semimajor axis of about 1500 AU with a tail out beyond 4000 AU. We constrained the inclination of the orbit with respect to the galactic plane, $i_b$, by calculating the location of the orbital pole\cite{chang29, heintz69, agati15} for each posterior sample from our fit. In particular, we used the equations on page 13 of Chang (1929\cite{chang29}), after correcting an error in the second equation on page 13 that should read $\sin{i}\sin{\Omega} = m\sin{M}$ (see Heintz 1969\cite{heintz69}). The probability distribution for $i_b$ is strongly peaked towards high inclinations, with the greatest probability at 90$^\circ$. At 68\% and 95\% confidence, $i_b$ must be greater than 60$^\circ$ and 41$^\circ$, respectively. Therefore, the galactic tide mechanism could plausibly operate in at least part of allowed orbital parameter space. }

\textit{Tidal dissipation during the giant phase:} Previously, Adams \& Bloch (2013\cite{Adams2013}) calculated the orbital evolution of exoplanets orbiting near expanding giant stars (see also Rasio et al. 1996\cite{rasiotout}). The orbits of these planets evolve due to two competing factors: mass loss (which drives orbits outwards) and tidal dissipation (which drives orbits inwards). Planets which orbit near an equilibrium radius where these two effects are nearly equal in strength can in some cases migrate inwards due to tidal evolution but avoid engulfment by the red giant host. This requires extremely finely-tuned initial parameters to have a chance of forming WD 1856 b's present-day configuration. Computing the exact location of this radius (which is likely somewhere around 1-2 AU) is difficult as the radius depends on the starting angular momentum, mass loss rate, dissipation coefficients, and other parameters that are difficult to constrain; however, it could be plausible that finely tuning the initial parameters of the planetary orbit and stellar properties could shrink the orbit of WD 1856 b to its current semimajor axis. 

\textit{Dynamical interactions near periastron:} If two planets happened to be scattered into close periastron passages at the same time and had a close scattering event near periastron, one planet could have been ejected, leaving the other planet in a short-period orbit around WD 1856. The likelihood of such an encounter is fairly low; events which can excite high eccentricities and close periastron distances are already rare (happening perhaps once in the lifetime of a white dwarf planetary system\cite{verasunpacking}), so the probability of two planets having close periastron passages simultaneously is even lower. Another related mechanism involves a proto-WD 1856 b with a massive moon (or a binary planet) on a highly eccentric orbit with a close periastron passage. The moon/binary companion could be ejected\cite{Payne2016} in a similar way to how hypervelocity stars are ejected binary members perturbed by the Galaxy's central black hole\cite{Bromley2012}, shedding enough orbital energy to leave WD 1856 b in a nearly circular orbit.  Again, this \bedit{mechanism} is \textit{a priori} unlikely, since we have yet to discover a binary planet. 

\textit{Partial Tidal Disruption:} If WD 1856 b reached a periastron distance slightly closer to WD 1856 than the Roche limit, it could have been partially tidally disrupted, losing enough mass to dissipate its orbital energy, while remaining at least somewhat intact\cite{faber}. This process has also been studied in the case of the tidal disruption of a star by a supermassive black hole\cite{Mainetti17}. If this process happened recently and material from the planet was still accreting onto the white dwarf, the elements might be visible in the planet's spectrum. This motivates more sensitive spectroscopy of WD 1856.

\subsection{Expected amplitude of spectral features in transmission}
Due to the small radius of the white dwarf host star, the spectral features expected from transmission spectroscopy are much larger than they would be around a main sequence star. We estimated the amplitude of spectral features as follows:

Traditionally the amplitude of spectral features in transmission is proportional to the annulus of the planet's terminator region\cite{kreidberg18}. However, that approximation does not apply to the case of a grazing transit where the star is smaller than the planet. To account for the grazing geometry for WD 1856, we assumed that the atmosphere covers a slice of the star with width equal to the stellar diameter and height equal to the scale height. In this case, the amplitude $A$ of spectral features is \begin{equation}\label{transmissionheight}
    A \approx \frac{2nH}{\pi R_\star}
\end{equation}
where $n$ is the number of scale heights typically crossed by atmospheric features (usually $n = 2$ for cloud-free gas giant exoplanets\cite{stevenson2016}) and $H$ is the atmospheric scale height:
\begin{equation}
    H = \frac{kT}{\mu g}
\end{equation}
where k is Boltzmann's constant, T is the planet's temperature, $\mu$ is the mean molecular weight in the atmosphere, and $g$ is the planet's surface gravity. To calculate the scale height, we assumed a solar composition atmosphere ($\mu$ = 2.3 amu) and assumed planet properties for two cases:

\begin{enumerate}
    \item Mp = 10 Mjup, T = 280 K (a reasonable internal temperature for an object of this mass)
    \item Mp = 1 Mjup, T = 165 K (the equilibrium temperature)
\end{enumerate}

For case 1, the scale height $H = 4$ km and the amplitude of spectral features is 0.1\%. For case 2, the scale height $H = 12$ km and the amplitude of spectral features is 0.7\%.

\bedit{We note that our assumption that the atmosphere covers a slice of the star with width equal to the stellar diameter is an approximation for nearly 50\% deep transits of planets that are much larger than their stars. A more general expression (valid for $\lvert 1 - Rp/R_\star \rvert < b < 1+ Rp/R_\star$) for the expected height of transmission features for grazing transits is \begin{equation}
    A \approx \frac{snH}{\pi R_\star}
\end{equation}
where
\begin{equation}
    s = 2\,\frac{R_p}{R_\star}\,\cos^{-1}\left(\frac{b^2 - 1 + \left(\frac{R_p}{R_\star}\right)^2}{2\,b\,\frac{R_p}{R_\star}}\right)
\end{equation}
For cases like WD 1856, where the planet is much larger than the star and blocks close to 50\% of the stellar disk, $s\approx2$, and the expression reduces to to Equation \ref{transmissionheight}. For WD 1856 b's particular transit parameters, $s = 2.004$. }

\subsection{Expected amplitude of Doppler boosting signal}
\bedit{WD 1856 b's mass could be measurable via small variations in the host star's brightness caused by Doppler boosting\cite{loebgaudi, vanKerkwijk10}. The semi-amplitude $A_b$ of the Doppler boosting signal is
\begin{equation}
    A_{b} = (3-\alpha)\frac{K}{c}
\end{equation}
where $K$ is radial velocity semiamplitude induced by the planet, $c$ is the speed of light, and $\alpha$ is the average logarithmic derivative of flux with respect to frequency. For a blackbody spectrum, $\alpha$ is approximately
\begin{equation}
 \alpha \simeq 3 - \frac{xe^x}{e^x-1}
\end{equation}
and \begin{equation}
    x = \frac{h\nu}{kT_{\rm eff}}
\end{equation}
where $h$ is Planck's constant, $\nu$ is the frequency of light in the observed bandpass, $k$ is Boltzmann's constant, and $T_{\rm eff}$ is the blackbody temperature. Assuming a mass of 14 \mjup\ for WD 1856 b, the Doppler boosting amplitude is about 50 parts per million (ppm) in the \TESS\ bandpass, about 100 ppm in blue optical light, and about 30 ppm in near infrared light around 1.5 microns.}

\bedit{It will be difficult to detect these signals because of WD 1856's intrinsic faintness and contamination from G 229-20 A/B. We fit the out-of-transit \TESS\ light curve (with a dilution correction applied) with a sine/cosine model and found a boosting semiamplitude of  $-770 \pm 1130$ ppm -- far too uncertain to detect an orbiting planet. If the \PLATO\ mission\cite{plato2} observes WD 1856 near the center of its field of view for two years, it may come close to a tentative detection of a 14 \mjup\ planet, depending on how much starlight from G 229-20 A/B contaminates WD 1856's aperture. With their large apertures and high spatial resolution, \JWST\ and \HST\ could detect the boosting signal, but the observations would be expensive. A 3$\sigma$ detection of a 14 \mjup\ planet would likely require $\gtrsim 10$ days of observations.}

\vspace{2cm}

\end{methods}

%\bibliographystyle{naturemagfixed}                                    
%\bibliography{refs}  

%% Here is the endmatter stuff: Supplementary Info, etc.
%% Use \item's to separate, default label is "Acknowledgements"

\begin{addendum}
\item  We thank Sebastian Lepine for providing the archival spectrum of G 229-20 A, and Perry Berlind and Jonathan Irwin for collecting and extracting velocities from the TRES spectrum. We thank Brice-Olivier Demory for helpful comments on the manuscript and Fred Rasio, Dimitri Veras, Peter Gao, Benjamin Kaiser, Willie Torres, Jonathan Irwin, JJ Hermes, Jason Eastman, Avi Shporer, and Keith Hawkins for helpful conversations. A.V.'s work was performed under contract with the California Institute of Technology (Caltech)/Jet Propulsion Laboratory (JPL) funded by NASA through the Sagan Fellowship Program executed by the NASA Exoplanet Science Institute.  I.J.M.C. acknowledges support from the NSF through grant AST-1824644, and from NASA through Caltech/JPL grant RSA-1610091. TD acknowledges support from MIT’s Kavli Institute as a Kavli postdoctoral fellow. D. D. acknowledges support from NASA through Caltech/JPL grant RSA-1006130 and through the TESS Guest Investigator Program Grant 80NSSC19K1727. S.B. acknowledges support from the Laboratory Directed Research and Development program of Los Alamos National Laboratory under project number 20190624PRD2. Resources supporting this work were provided by the NASA High-End Computing (HEC) Program through the NASA Advanced Supercomputing (NAS) Division at Ames Research Center for the production of the SPOC data products. This work is partially based on observations made with the Nordic Optical Telescope, operated by the Nordic Optical Telescope Scientific Association at the Observatorio del Roque de los Muchachos, La Palma, Spain, of the Instituto de Astrofisica de Canarias. This article is partly based on observations made with the MuSCAT2 instrument, developed by ABC, at Telescopio Carlos S\'anchez operated on the island of Tenerife by the IAC in the Spanish Observatorio del Teide. This work is partly supported by JSPS KAKENHI Grant Numbers JP17H04574, JP18H01265 and JP18H05439, and JST PRESTO Grant Number JPMJPR1775. This research has made use of NASA's Astrophysics Data System, the NASA Exoplanet Archive, which is operated by the California Institute of Technology, under contract with the National Aeronautics and Space Administration under the Exoplanet Exploration Program, and the SIMBAD database, operated at CDS, Strasbourg, France. This work is based in part on observations made with the \Spitzer\ Space Telescope, which is operated by the Jet Propulsion Laboratory, California Institute of Technology under a contract with NASA. This work is partially based on observations obtained at the international Gemini Observatory, a program of NOIRLab, which is managed by the Association of Universities for Research in Astronomy (AURA) under a cooperative agreement with the National Science Foundation. on behalf of the Gemini Observatory partnership: the National Science Foundation (United States), National Research Council (Canada), Agencia Nacional de Investigaci\'{o}n y Desarrollo (Chile), Ministerio de Ciencia, Tecnolog\'{i}a e Innovaci\'{o}n (Argentina), Minist\'{e}rio da Ci\^{e}ncia, Tecnologia, Inova\c{c}\~{o}es e Comunica\c{c}\~{o}es (Brazil), and Korea Astronomy and Space Science Institute (Republic of Korea). The authors wish to recognize and acknowledge the very significant cultural role and reverence that the summit of Maunakea has always had within the indigenous Hawaiian community.  We are most fortunate to have the opportunity to conduct observations from this mountain.
 
\item[Contributions] A.V. led the \TESS\ proposals, identified the planet candidate, organized observations, performed the transit and flux limit analysis, and wrote most of the manuscript. S.A.R helped organize observations, performed independent data analysis, and wrote portions of the manuscript. S.X. helped organize observations, obtained and analyzed the Gemini data, measured fluxes from the \Spitzer\ data, and helped guide the strategy of the manuscript. I.J.M.C, La.K, V.G., B.B., D.B., J.L.C., D.D., C.D., S.R.K., Fa.M., and L.Y acquired and produced a light curve from the \Spitzer\ data. S.A.R., J.C.B., L.N., B.Z., F.C.A., and J.J.L investigated the formation of the WD 1856 system. B.G, Fe.M, T.G.K., E.P., H.P., A.F., and N.N. acquired follow-up photometry. S.B., P.D., and K.G.S. determined the parameters of the white dwarf, while A.W.M. and E.R.N. studied the M-dwarf companions. C.M., G.Z., W.R.B., R.T., L.A.B., A.E.D., and A.I.H. acquired spectra of the white dwarf and/or M-dwarf companions. B.M.,  K.H., and T.D. performed an independent analysis analysis of the \TESS\ data, and J.A.L. performed an independent analysis of the white dwarf SED. C.V.M. provided valuable expertise on brown dwarf models, and Li.K. investigated the system's implications. L.A.P. determined parameters for the binary M-dwarf orbits and white dwarf/M-dwarf orbits, and A.C. investigated the system's galactic kinematics.  G.R.R., R.K.V., D.W.L., S.S., J.N.W., J.M.J., D.A.C., K.A.C., K.D.C., J.D., A.G., N.M.G., C.X.H., J.P., M.E.R., and J.C.S. are members of the \TESS\ mission team. 

\item[Data Availability] We provide all reduced light curves and spectra as supplementary data products to the manuscript. 

 \item[Competing Interests] The authors declare that they have no competing financial interests.
 
 \item[Correspondence] Correspondence and requests for materials should be addressed to A.V.~(email: avanderburg@wisc.edu).
 
 \item[Code availability] Much of the code used to produce these results is publicly available and linked throughout the paper. We wrote custom software to analyze the data collected in this project. Though this code was not written with distribution in mind, we will make our custom analysis code available online via github for transparency. 
 
\end{addendum}

%%
%% TABLES
%%
%% If there are any tables, put them here.
%%

\end{document}